\documentclass[review]{elsarticle}

\usepackage{lineno,hyperref}
\usepackage{amsmath}
\usepackage{array}
\usepackage{tabularx}
\usepackage{booktabs}
\usepackage{mathtools}
\usepackage{natbib}
\usepackage{hyperref}
\usepackage{hypernat}
\modulolinenumbers[5]

\makeatletter
\newcommand{\thickhline}{%
    \noalign {\ifnum 0=`}\fi \hrule height 1pt
    \futurelet \reserved@a \@xhline
}
\makeatother

\journal{Journal of Computational Science}

\begin{document}

\begin{frontmatter}

\title{DSMC-LBM mapping scheme \\for rarefied and non-rarefied gas flows}

%%% Group authors per affiliation:

\author[TUe] {G. Di Staso\corref{cor1}}
\ead{g.di.staso@tue.nl, +31 40 247 21 43}

\author[TUe] {H.J.H. Clercx}
\ead{h.j.h.clercx@tue.nl}

\author[CNR] {S. Succi}
\ead{succi@iac.cnr.it}

\author[TUe,TUemath,CNR] {F. Toschi}
\ead{f.toschi@tue.nl}

\cortext[cor1]{Corresponding author}

\address[TUe]{Department of Applied Physics, Eindhoven University of Technology,\\
Den Dolech 2, 5600 MB, Eindhoven, Netherlands}

\address[TUemath]{Department of Mathematics and Computer Science, Eindhoven University of Technology,\\
Den Dolech 2, 5600 MB, Eindhoven, Netherlands}

\address[CNR]{Istituto per le Applicazioni del Calcolo, Consiglio Nazionale delle Ricerche,\\
Via dei Taurini 19, 00185, Rome, Italy}

\begin{abstract}
We present the formulation of a kinetic mapping scheme between the Direct Simulation
Monte Carlo (DSMC) and the Lattice Boltzmann Method (LBM) which is at the basis
of the hybrid model used to couple the two methods in view of efficiently and
accurately simulate isothermal flows characterized by variable rarefaction effects.
Owing to the kinetic nature of the LBM, the procedure we propose
ensures to accurately couple DSMC and LBM at a larger Kn number than usually done in 
traditional hybrid DSMC---Navier-Stokes equation models.
We show the main steps of the mapping algorithm and illustrate details 
of the implementation. Good agreement is found between the moments of 
the single particle distribution function
as obtained from the mapping scheme and from independent LBM or DSMC simulations 
at the grid nodes where the coupling is imposed. 
We also show results on the application of the hybrid scheme based on a 
simpler mapping scheme for plane Poiseuille flow at finite Kn number. 
Potential gains in the computational efficiency assured by the application 
of the coupling scheme are estimated for the same flow.
\end{abstract}

\begin{keyword}
kinetic theory \sep Grad's moments method \sep non-equilibrium effects \sep rarefied 
gas flows \sep hybrid method
\MSC[2010] 00-01\sep  99-00
\end{keyword}

\end{frontmatter}

\linenumbers

\section{Introduction}

Research in gas flows characterized by a large range of scales and
by disparate levels of non-equilibrium effects poses a challenge to 
statistical physics modelling and rises interest in industry
for simulating flows in micro-, nano-electromechanical systems
and in material processing tools \cite{Ho1998,Reese2003,Karnadiakis2005}.
The extent of the departure of a flow from the equilibrium state 
is traditionally measured in terms of the Knudsen number:
\begin{equation}
\textrm{Kn}=\frac{\lambda}{\ell} \approx \frac{\lambda}{Q} \left|\frac{dQ}{d\ell}\right|,
\end{equation}
where $\lambda$ is the gas mean free path, $\ell$ is the smallest
hydrodynamic characteristic scale and $Q$ is a fluid dynamic quantity
of interest such as the gas pressure, velocity, temperature
\cite{Boyd1995}.
According to the Knudsen number, the gas flows can be classified
into the hydrodynamic (Kn$<0.01$), slip (0.01$<$Kn$<$0.1),
transition (0.1$<$Kn$<$10) and free molecular regime (Kn$>$10).
The kinetic description of gases based on the Boltzmann equation,
valid at any Kn, allows to cover flow conditions from
the very rarefied to the hydrodynamic limit \cite{Huang1987}.
The two limits, rarefied and continuum, have traditionally been
studied numerically by approximating the Boltzmann equation via
the Direct Simulation Monte Carlo (DSMC) 
\cite{Bird1994} 
or by solving the Navier-Stokes equations which can be derived from
truncation at first order of the Chapman-Enskog procedure \cite{ChapmanCowling1970}.
While the DSMC method is particularly suited to rarefied
gas flow (transitional regime), its computational costs make 
it unpractical to study hydrodynamic flows
\cite{Reese2003}.
Conversely, the continuum description of the flow provided by
solving the Navier-Stokes equations and applying the no-slip boundary condition
is not accurate whenever Kn$>$0.01
\cite{Garcia1999}.
Corrections to the boundary conditions of Navier-Stokes equations such as
to reproduce the velocity slip and temperature jump at the gas-surface
interface in case of slip flow regime are often not accurate and
may also predict incorrect qualitative behavior of the flow
\cite{MalekMansours1997,Zhang2012}.
Moreover, the derivation of extended hydrodynamic equations employing 
higher-order Chapman-Enskog approximations (Burnett and super-Burnett 
equations) have showed limited success
\cite{Garcia1999}.
Alternatively, macroscopic transport equations can be originated 
from moments expansion methods such as the Grad's method
\cite{Grad1949a,Struchtrup2005}.
However, difficulties in imposing boundary conditions for those 
moments without a clear physical meaning, as well complexity
in the resulting systems of equations prevent the application
of the method for the simulation of flows of industrial interest.\\
It is therefore evident that whenever the flow presents a large range of
Kn, due to the current computational and modelling limitations of the
available methods, a multiscale hybrid model has to be used.\\
When dealing with multiscale models, \emph{domain decomposition techniques}
represent the most natural way to handle the problem.
Within this approach, the domain is decomposed according to a continuum breakdown 
parameter between regions where continuum-level macroscopic
equations (either Euler or Navier-Stokes equations) are valid and 
regions where substantial non-equilibrium effects are present and 
kinetic methods, typically DSMC, are needed (see Refs. 
\cite{Garcia1999,Roveda1998,Wijesinghe2004,Wu2006,Schwatzentruber2007,Kessler2010,Pantazis2014,Farber2015}). 
Then a special treatment is imposed to couple
the flow fields in the areas of overlap between the different
regions, e.g. \cite{Bourgat1996,LeTallec1997,Garcia1998}.\\
For completeness, the domain decomposition technique is not 
the only method adopted in the literature as alternative approaches 
are proposed. For example in \cite{Almohssen2007}, the Boltzmann
equation is solved for a short period of time to obtain the rate of
change of the average flow variables which are then used to update the continuum-level
velocity field. In \cite{Degond2006}, instead, macroscopic 
equations are modified so to include effects due to kinetic contributions 
which take into account perturbations from the equilibrium state of the 
velocity distribution.\\
The approach that we introduce here follows the domain
decomposition technique as commonly done in models proposed in literature but it
departs from those as
%, for the first time, 
the flow at the continuum level and at moderate rarefied conditions
is simulated with the Lattice Boltzmann Method (LBM).\\
Moreover, since it has been largely demonstrated that LBM, due
to its intrinsic kinetic nature, is an accurate and efficient
numerical solver not only for flows at Navier-Stokes description
level but also for flows at finite Kn number (see Refs.
\cite{Succi2002,Toschi2005,Sbragaglia2005,Sbragaglia2006,Zhang2006,Ansumali2007,Niu2007,Tanga2008,Menga2011,Mengb2011,Reis2012,Liu2013,Tao2015}), 
the present model has the advantage, over the other
hybrid methods which use traditional Navier-Stokes solvers,
that the need for using the computationally expensive DSMC
solver can be postponed to larger values of Kn.
This is equivalent to say that the size of the domain where
DSMC solver is still needed can be significantly reduced,
thus improving the overall computational efficiency of the
simulation.\\
In this work we principally focus on the most delicate aspect of
any hybrid coupling model, i.e. the two-way extraction and
transfer of information at the interface between the two
numerical methods.
The \emph{mapping} schemes we developed, in fact, allow to pass
from DSMC to LBM domains and vice versa correctly transferring 
also the non-equilibrium information. The \emph{amount} of
non-equilibrium information that can be passed is then essentially 
determined by the LB model and in particular by the chosen set
of discrete velocities and the isotropy conditions the set is
able to fulfill. \\
Simulations performed to validate the mapping scheme show that
an accurate transfer of information is achievable for flows up
to Kn=0.25 for a 39-points Gauss-Hermite quadrature with sixth-order
isotropy (D3Q39).\\
Finally, to check functionality of the DSMC-LBM hybrid model and 
assess its computational efficiency, tests, based on a simpler
mapping scheme, are also performed showing, for the particular
simulated flow, a significant speed-up with respect to a full 
DSMC simulation.

\section{Mapping schemes}\label{sec:remapping}
Since both LB and DSMC are widely documented in the literature, only a
few basic aspects are discussed in this paper.
For an exhaustive treatment about DSMC and LBM methods, the reader
should refer to \cite{Bird1994} and \cite{Succi2001}.\\
Both methods aim to determine the fluid motion as described by the
Boltzmann equation. The main feature which clearly distinguishes
the LBM from the DSMC, is the reduction of the degrees of
freedom of the velocity space. In fact in LBM particles at each lattice site $\mathbf{x}$ can only
propagate along a finite number of directions with an assigned speed $\boldsymbol\xi_a$,
while in DSMC the velocity space is not constrained to a set of discrete velocities.\\
Before introducing the mapping scheme between the DSMC and LBM, we note
that in order to quantitatively reproduce DSMC solutions for finite-Kn
number flows, the LB model needs three basic \emph{ingredients}:
\begin{enumerate}[1.]
\item kinetic boundary conditions, \cite{ZhangJF2011,Verhaeghe2009,Guo2014,Ansumali2002,Chai2008,Tao2015};
\item higher-order lattice (HOL), \cite{Shan2006,Zhang2006};
\item regularization procedure, \cite{Montessori2014,Zhang2006}.
\end{enumerate}
The main idea at the basis of the \emph{mapping scheme} is that the single particle
distribution function $f(\mathbf{x},\boldsymbol\xi,t)$ can be expanded in
terms of the dimensionless Hermite orthonormal polynomials, $\mathcal{H}(\boldsymbol\xi)$,
in the velocity space $\boldsymbol\xi$ as \cite{Grad1949a,Struchtrup2005,Shan2006}:
\begin{equation}
f(\mathbf{x},\boldsymbol\xi,t) = \omega(\boldsymbol\xi) \sum_{n=0}^{\infty} \frac{1}{n!} \mathbf{a}^{(n)}(\mathbf{x},t)\mathcal{H}^{(n)}(\boldsymbol\xi),\label{eq:series}
\end{equation}
where $\omega(\boldsymbol\xi)$ is the weight function associated with the Hermite
polynomials, and $\mathbf{a}^{(n)}$ are the rank-$n$ tensors representing the
dimensionless expansion coefficients defined as:
\begin{equation}
\mathbf{a}^{(n)}=\int f(\mathbf{x},\boldsymbol\xi,t) \mathcal{H}^{(n)}(\boldsymbol\xi) d\boldsymbol\xi. \label{eq:nordercoefficients}
\end{equation}
The first coefficients of the series, due to the definition of the Hermite
polynomials, can be identified as the hydrodynamic moments (or a combination
of those) of the distribution $f(\mathbf{x},\boldsymbol\xi,t)$:
\begin{equation}
\small
\mathbf{a}^{(0)}=\int f(\mathbf{x},\boldsymbol\xi,t) \mathcal{H}^{(0)}(\boldsymbol\xi) d\boldsymbol\xi = \int f(\mathbf{x},\boldsymbol\xi,t) d\boldsymbol\xi = \rho, \label{eq:density}
\end{equation}
\begin{equation}
\small
\mathbf{a}^{(1)}=\int f(\mathbf{x},\boldsymbol\xi,t) \mathcal{H}^{(1)}(\boldsymbol\xi) d\boldsymbol\xi =\\
\int f(\mathbf{x},\boldsymbol\xi,t) \boldsymbol\xi d\boldsymbol\xi = \rho \mathbf{u} \label{eq:momentum}
\end{equation}
\normalsize
and analogously for higher-order coefficients.\\
Due to the orthonormality of the Hermite polynomials,
\begin{equation}
\small
f(\mathbf{x},\boldsymbol\xi,t) \approx f^N(\mathbf{x},\boldsymbol\xi,t) =\omega(\boldsymbol\xi) \sum_{n=0}^{N} \frac{1}{n!} \mathbf{a}^{(n)}(\mathbf{x},t)\mathcal{H}^{(n)}(\boldsymbol\xi)\label{eq:seriestrunc}
\end{equation}
and $f^N(\mathbf{x},\boldsymbol\xi,t)$ has the same leading $N$ velocity moments as the 
complete $f(\mathbf{x},\boldsymbol\xi,t)$.\\
It is possible now to describe the two mapping procedures:
\begin{itemize}
\item[-] the DSMC2LB (or \emph{projection}) step that allows to \emph{project}
the DSMC hydrodynamic variables (fine level of description) onto the
LBM discrete distributions (coarse level of description);
\item[-] the LB2DSMC (or \emph{reconstruction}) step that allows to 
\emph{reconstruct} from the LBM discrete distributions (coarse level),
the continuous, truncated, distribution function (fine level) from which the
velocities of the DSMC particles can be sampled, e.g. via acceptance/rejection
method.
\end{itemize}
It has to be noted that the following procedures can be extended to any 
suitable LB stencil whose discrete speeds are actually abscissae
of a Gauss-Hermite quadrature.
\subsection{DSMC2LB mapping scheme}\label{sec:dsmc2lb}
Firstly, we present the DSMC2LB projection step.
In correspondence with the DSMC cells/LBM nodes
where the coupling occurs, the cumulative averages of the DSMC hydrodynamic variables,
properly scaled (see \ref{sec:appendix} on how to perform such scaling), are used 
to compute the coefficients $\mathbf{a}_{\textrm{DSMC}}^{(n)}$ of the 
truncated distribution $f^N_{\textrm{DSMC}}(\mathbf{x},\boldsymbol\xi,t)$ in Eq. (\ref{eq:seriestrunc}).\\
We now take advantage of the fact that the distribution $f^N_{\textrm{DSMC}}(\mathbf{x},\boldsymbol\xi,t)$
can be completely and uniquely determined by its values at a set of discrete velocities
and, if the Gauss-Hermite quadrature is used, then the coefficients $\mathbf{a}_{\textrm{DSMC}}^{(n)}$
can be expressed as:
\begin{equation}
\begin{split}
\small
\mathbf{a}_{\textrm{DSMC}}^{(n)}=\int f^N_{\textrm{DSMC}}(\mathbf{x},\boldsymbol\xi,t) \mathcal{H}^{(n)}(\boldsymbol\xi) d\boldsymbol\xi =\\
\sum_{a=0}^{d-1} \frac{w_a}{\omega(\boldsymbol\xi_a)} f^N_{\textrm{DSMC}}(\mathbf{x},\boldsymbol\xi_a,t) \mathcal{H}^{(n)}(\boldsymbol\xi_a), \label{eq:coefficients}
\end{split}
\end{equation}
where $w_a$ and $\boldsymbol\xi_a$ are the weights and abscissae of a Gauss-Hermite
quadrature of algebraic precision of degree $\geq 2N$, and $d$
is the total number of discrete velocities of the quadrature.\\
The definitions of the first two hydrodynamic moments in the LBM are:
\begin{equation}
\rho=\sum_{a} f_a,\:\:\:\: \rho \mathbf{u}= \sum_a f_a \boldsymbol\xi_a. \label{eq:lbmmoments}
\end{equation}
Comparing Eq. (\ref{eq:coefficients}) with Eq. (\ref{eq:lbmmoments}) and
recalling the definitions of the Hermite polynomials $\mathcal{H}^{(n)}$ 
and that the coefficients $\mathbf{a}^{(n)}$ are the velocity moments of the 
$f^N(\mathbf{x},\boldsymbol\xi,t)$, or a proper combination of those, 
it is immediate to see that the discrete distributions are the scaled values of the continuous 
distribution function evaluated at the Gauss-Hermite quadrature abscissae $\boldsymbol\xi_a$:
\begin{equation}
f_{\textrm{DSMC2LB},a}(\mathbf{x},t) = \frac{w_a f^N_{\textrm{DSMC}}(\mathbf{x},\boldsymbol\xi_a,t)}{\omega(\boldsymbol\xi_a)}. \label{eq:discretefunctions}
\end{equation}
Therefore, once the $f^N_{\textrm{DSMC}}(\mathbf{x},\boldsymbol\xi,t)$ is built from the DSMC
hydrodynamic moments and evaluated at the quadrature abscissae, 
$f^N_{\textrm{DSMC}}(\mathbf{x},\boldsymbol\xi_a,t)$,
the discrete (non-equilibrium) distributions to be supplemented to the LBM solver
at the coupling nodes can be computed from Eq. (\ref{eq:discretefunctions}).
\subsection{LB2DSMC mapping scheme}\label{sec:lb2dsmc}
\begin{figure}
\begin{center}
\includegraphics[width=0.6\textwidth]{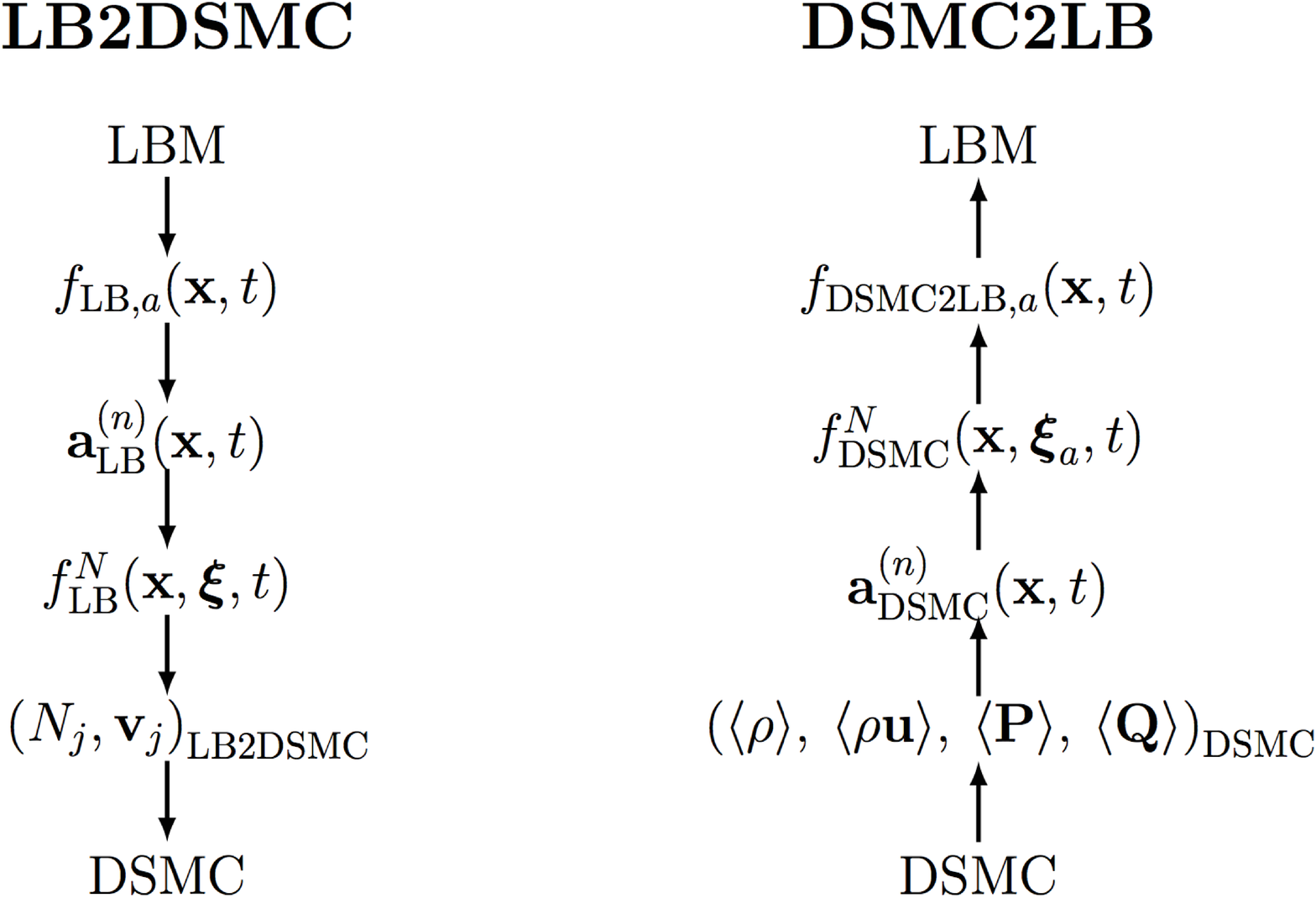}
\caption{Schematic showing the main steps of the top-down LB2DSMC reconstruction (left) 
and bottom-up DSMC2LB projection (right) mapping schemes as described in Section
\ref{sec:remapping}. $\mathbf{P}$ and $\mathbf{Q}$ represent the second 
and third order momentum flux tensors, respectively. Symbol $\langle\: \rangle$
represents the cumulative average measurements of hydrodynamic moments 
from the DSMC solver.}\label{fig:schematic_remapping}
\end{center}
\end{figure}

The inverse reconstruction step (LB2DSMC) requires that at the LBM lattice nodes/DSMC cells
where the coupling occurs, the velocities of the DSMC particles are sampled from a continuous
distribution function.\\
At those lattice sites, the LBM discrete non-equilibrium functions $f_{\textrm{LB},a}$, are used
to compute the coefficients of the expansion in Eq. (\ref{eq:seriestrunc}):
\begin{equation}
\mathbf{a}^{(n)}_{\textrm{LB}}=\sum_{a=0}^{d-1} f_{\textrm{LB},a}(\mathbf{x},t) \mathcal{H}^{(n)}(\boldsymbol\xi_a) \label{eq:coefficients_LBM}
\end{equation}
These allow to build the continuous truncated distribution $f^N_{\textrm{LB}}(\mathbf{x},\boldsymbol\xi,t)$.
To generate the velocities of the DSMC particles, the distribution should be sampled.\\
Several algorithms can be employed to this aim. We chose to adopt an \emph{acceptance/rejection}
algorithm similar to the one presented in \cite{Garcia1998}. However, while in \cite{Garcia1998}
a Chapman-Enskog distribution was sampled, in the present case a Grad's distribution has to be sampled but,
nonetheless, most of the steps presented there can be used here.\\
The Grad's velocity distribution, truncated up to order $N$, can be written as
\begin{equation}
g^{N}(\mathbf{x},\boldsymbol\xi,t) = g^{(0)}(\boldsymbol\xi)G(\mathbf{x},\boldsymbol\xi,t)\label{eq:sampledistr}
\end{equation}
where $g^{(0)}(\boldsymbol\xi)$ is the weight function associated with the Hermite polynomials
\begin{equation}
g^{(0)}(\boldsymbol\xi) = \omega(\boldsymbol\xi) = \frac{1}{(2 \pi)^{D/2}} \exp \left(-\frac{\xi^2}{2}\right)\label{eq:globalMax}
\end{equation}
with $D$ being the dimensionality of the flow problem.
Eq. (\ref{eq:globalMax}) represents also a global Maxwell-Boltzmann distribution at thermodynamic equilibrium
(here we set a constant temperature $T=1$ as we are interested in isothermal flows).\\
At thermodynamic equilibrium $G(\mathbf{x},\boldsymbol\xi,t)=1$, while away from that
condition, it can be expressed as:
\begin{equation}
\small
\begin{split}
G(\mathbf{x},\boldsymbol\xi,t) = 1 + \frac{1}{2!}\mathbf{a}^{(2)}_{\textrm{LB}} \mathcal{H}^{(2)}(\boldsymbol\xi) \\
+ \frac{1}{3!} \mathbf{a}^{(3)}_{\textrm{LB}} \mathcal{H}^{(3)}(\boldsymbol\xi) + \dots + \frac{1}{N!}\mathbf{a}^{(N)}_{\textrm{LB}} \mathcal{H}^{(N)}(\boldsymbol\xi)\label{eq:distr_outeq}
\end{split}
\end{equation}
The steps followed in the generation of the velocities of DSMC particles are 
outlined in Table \ref{tab:sampling}.
\begin{table}[ht!]
\centering
\begin{tabular}{ m{7cm} }
\hline
\hline
\small \textbf{Sampling acceptance/rejection algorithm for the Grad's distribution LB2DSMC}\\
\hline
\small
1. Compute the coefficients 
\begin{equation} \small a_{\textrm{LB},ij}^{(2)}=\sum_a f_{\textrm{LB},a} (\xi_{a,i}\xi_{a,j} - \delta_{ij}) \end{equation}
and similarly for the higher-order ones\\
\hline
\small 2. Find \begin{equation} \small M \equiv \max \left(\left|a_{\textrm{LB},ij}^{(2)} \right|, \left|a_{\textrm{LB},ijk}^{(3)} \right|,\dots, \left|a_{\textrm{LB},ijk\dots}^{(N)} \right|\right)\end{equation}\\
\hline
\small 3. Set the parameter \begin{equation}C = 1+ 30 M \label{eq:C} \end{equation}\\
\hline
\small 4. Sample a try particle velocity $\boldsymbol\xi_{\textrm{try}}$ from the Maxwell-Boltzmann distribution
$g^{(0)}(\boldsymbol\xi)$ using e.g. the Box-M\"{u}ller transformation method\\
\hline
\small 5. Accept the $\boldsymbol\xi_{\textrm{try}}$ if 
$C \mathcal{R} \leq G(\mathbf{x},\boldsymbol\xi_{\textrm{try}},t)$
with $\mathcal{R}$ a uniform deviate in the interval $[0,1)$ otherwise reject it and go back to step 4\\
\hline
\small 6. Generate the DSMC particle velocity as \begin{equation}\mathbf{v}_{j,\textrm{LB2DSMC}} = \left(\frac{2k_B T_{\textrm{DSMC}}}{m_{\textrm{DSMC}}}\right)^{1/2} \boldsymbol\xi_{\textrm{try}} + \mathbf{u}_{\textrm{LB}} \label{eq:step6} \end{equation}\\
\hline
\hline
\end{tabular}
\normalsize
\caption{Steps of the sampling acceptance/rejection algorithm for the LB2DSMC reconstruction
mapping scheme used to generate the velocities of DSMC particles from LBM data.}\label{tab:sampling}
\end{table}
Some comments on those steps.
The acceptance/rejection method needs to define an envelope
function $\gamma(\boldsymbol\xi)$ such that 
$\gamma(\boldsymbol\xi) \geq g(\boldsymbol\xi)$ for any $\boldsymbol\xi$.
In step 3, an amplitude parameter $C$ is set. In this way it 
is guaranteed that the function $\gamma(\boldsymbol\xi) = C g^{(0)}(\boldsymbol\xi)$
envelops most of the Grad's distribution function below it.
The larger this parameter, the less probable is the chance that
$G(\mathbf{x},\boldsymbol\xi,t)$ is larger than the envelop function,
but at the same time, the smaller the efficiency of the sampling method
since the efficiency is equal to $1/C$.
In step 6, the particle velocity is generated as the sum
of the thermal velocity and of the local fluid velocity. In Eq. (\ref{eq:step6}),
the thermal velocity is determined according to the temperature value
and to the molecular mass of the gas as set in the DSMC simulation.\\
Apart from the velocity, also the number of the DSMC particles, $N_{\textrm{LB2DSMC}}$,
must be set in order to guarantee conservation of mass at the coupling sites
so that the density from LBM and the density from DSMC, appropriately scaled, match
with each other.\\
In Figure \ref{fig:schematic_remapping}, the schematic showing the main 
steps involved in both the mapping schemes is drawn.\\
It is interesting to try to identify sources of inaccuracy in the proposed 
mapping scheme. In the reconstruction and projection steps, in fact, some
information is inevitably lost. In particular, in the LB2DSMC reconstruction
step, the truncated distribution, $f_{\textrm{LB}}^N \left(\mathbf{x},\boldsymbol\xi,t\right)$,
is derived from the discrete distributions, $f_{\textrm{LB},a}(\mathbf{x},t)$.
This truncated distribution is such that only the first $N$ moments are the
same as those of the non-truncated continuous distribution $f\left(\mathbf{x},\boldsymbol\xi,t\right)$,
with the value of $N$ essentially depending on the particular quadrature used. The moments of order higher
than $N$, in fact, will not be the same as those of the original continuous distribution.
This, in turn, reflects in the fact that the DSMC particles will be given a velocity
which is sampled from a distribution which accurately recovers up to the first $N$ moments.
If, then, the sampling process were perfectly able to sample the velocity distribution 
$f_{\textrm{LB}}^N \left(\mathbf{x},\boldsymbol\xi,t\right)$, then also the moments 
computed from the velocities of the particles would be perfectly reproduced in the
limit of an infinite number of independent samples. However, since only a finite number
of samples can be obtained, measurements of moments will be affected by statistical noise
which will be also present in the discrete distribution functions $f_{\textrm{DSMC2LB},a}(\mathbf{x},t)$.\\
Analogously, in the DSMC2LB projection step, the loss of information derives from the 
fact that only the first $N$ moments are used to evaluate the truncated discrete distributions
$f_{\textrm{DSMC2LB},a}(\mathbf{x},t)$, while, in principle, the DSMC solution possesses information
on all the moments up to $N\to \infty$.
The truncation, again, is performed according to the algebraic degree of accuracy of the
particular LB quadrature. To be more precise, this does not imply that moments of order larger
than $N$ cannot be evaluated but it means that they are not accurately computed.
If the so found discrete distributions were used to build a continuous distribution
from which to sample the velocities of the DSMC particles, then the source of inaccuracy
would be mainly related to the acceptance/rejection algorithm and in particular on the
choice of parameter $C$ in Eq. (\ref{eq:C}) which determines the extension of the envelope
function $\gamma(\boldsymbol\xi)$.

\section{Numerical results}

\subsection{Comparison between DSMC and LBM data}\label{subsec:comparison}
To understand and determine the extent of the overlap region where both DSMC and LBM provide
comparable accuracy in simulating rarefied gas flows, we performed independent 
force-driven plane Poiseuille flow simulations with two parallel plates 
at $x=0$ and $x=H$ and compared results obtained 
from D3Q19 and D3Q39 LB models with DSMC data. 
Even if the flow is strictly a monodimensional flow, we used 3D solvers 
since our final aim is to be able to simulate more complex flows.
This choice reflects in the fact that double periodic boundary conditions 
are imposed along the $y-$ and $z-$ directions.\\
Tests are performed at different Kn number, based on channel height, 
while keeping constant the Ma number, based on the flow centerline velocity, 
$u_{max}$: Ma = $u_{max}/c_s=0.1$.
The Ma number is set to such a value to guarantee that the lattice equilibria in LBM,
expressed as a second-order (D3Q19) or a third-order (D3Q39) expansion in Ma number of
the local Maxwellian, are positive defined, but it is still sufficiently high to prevent DSMC 
simulations from becoming impractically computationally expensive.\\
In the BGK-LBM simulations, we set the flow Kn number imposing the relaxation time
$\tau$ according to the relation \cite{Tangs2005,Tangb2008}:
\begin{equation}
\tau = \sqrt{\frac{\pi}{8}} \: \frac{c}{c_s} \: \textrm{Kn}\textrm{H} + 0.5 \label{eq:tau}
\end{equation}
where $c/c_s$ is equal to $\sqrt{3}$ for the D3Q19 model and to $\sqrt{3/2}$ for the
D3Q39 model, and H is the number of lattice sites along the channel height.
Once Kn and H are set, $\tau$ is also set. For both D3Q19 and D3Q39 models,
kinetic boundary conditions and regularization procedure are applied.\\
In the DSMC simulations, we set the Kn number imposing the height of the channel, H,
and the mean free path $\lambda$. To set $\lambda$, a proper number density $n$ and 
a collision model should be defined. In the case of Hard Sphere (HS) model, the
relation between $\lambda$ and $n$ (at equilibrium) is given by \cite{Bird1994}:
\begin{equation}
\lambda = \frac{1}{\sqrt{2} \pi d_{\textrm{ref}}^2 n} \label{eq:MFP}
\end{equation}
where $d_{\textrm{ref}}$ is a reference molecular diameter. The determination of $\lambda$
from Eq. (\ref{eq:MFP}) and estimates on the molecular speed allows to define
the space and time discretizations.\\
Once the number of cells along the channel height is determined from DSMC 
parameters, an equal number of lattice sites is imposed in the LBM simulation
so that the cells centers in DSMC and the LBM lattice sites overlap.
\begin{figure}
\begin{center}
\includegraphics[width=0.75\textwidth]{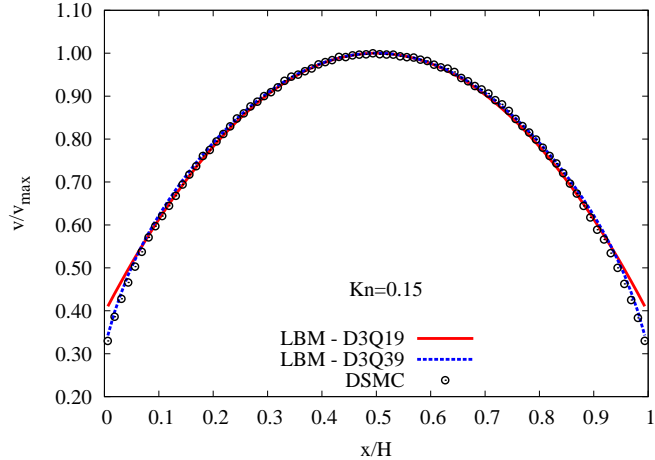}
\caption{The velocity profiles of the planar force-driven
Poiseuille flow for Kn=0.15. The LBM results with both the D3Q19 
and D3Q39 models are compared with the DSMC solution. For both
the LB models the regularization procedure is applied. Fully
diffuse reflection is imposed at the walls, $x/H=0$ and $x/H=1$,
for both the LBM and DSMC simulations.} \label{fig:profiles}
\end{center}
\end{figure}

\begin{figure}
\begin{center}
\includegraphics[width=0.55\textwidth]{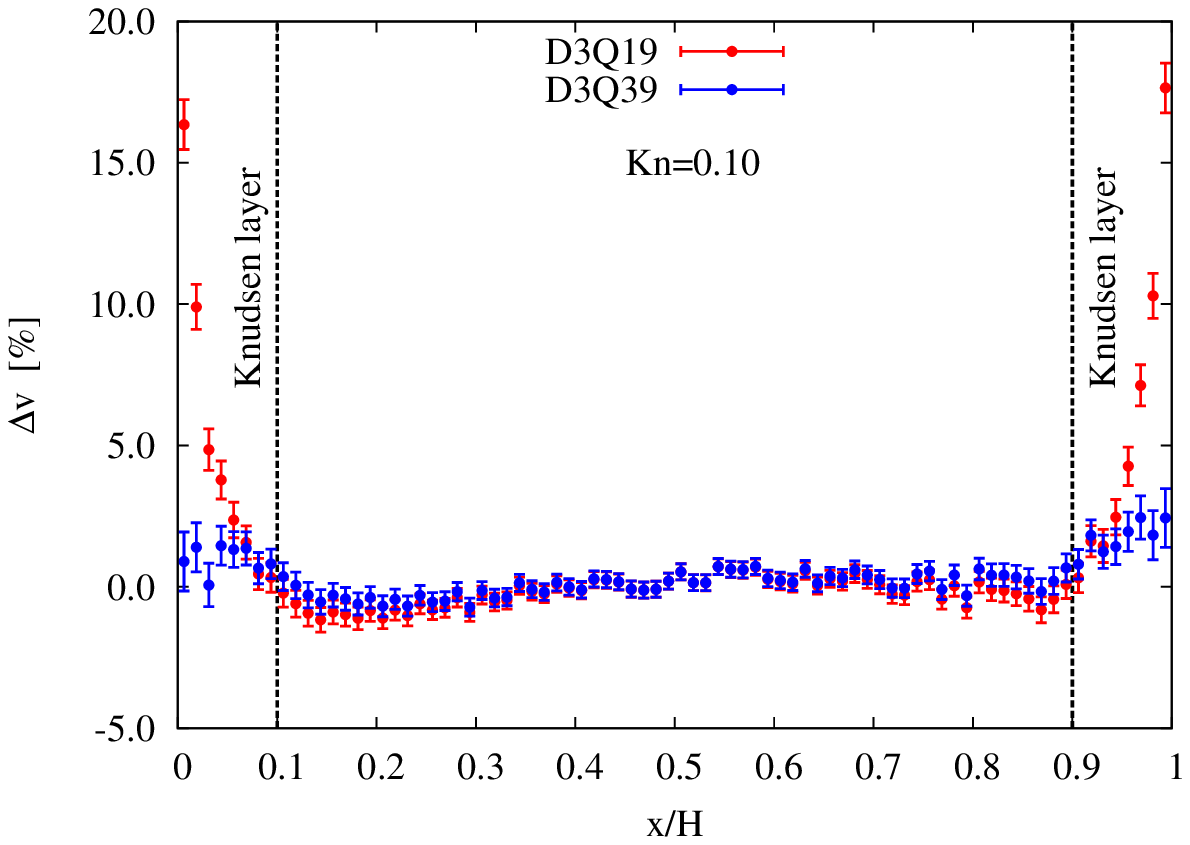}
\includegraphics[width=0.55\textwidth]{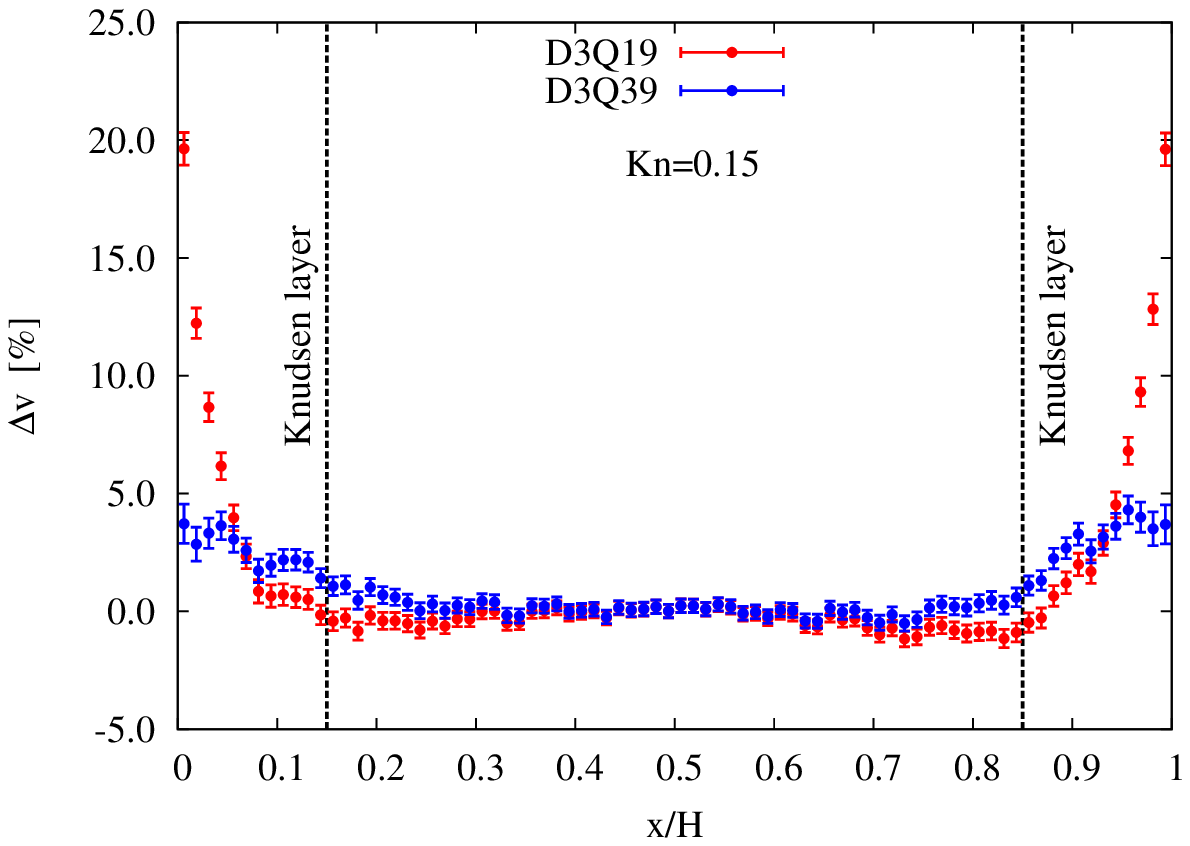}
\includegraphics[width=0.55\textwidth]{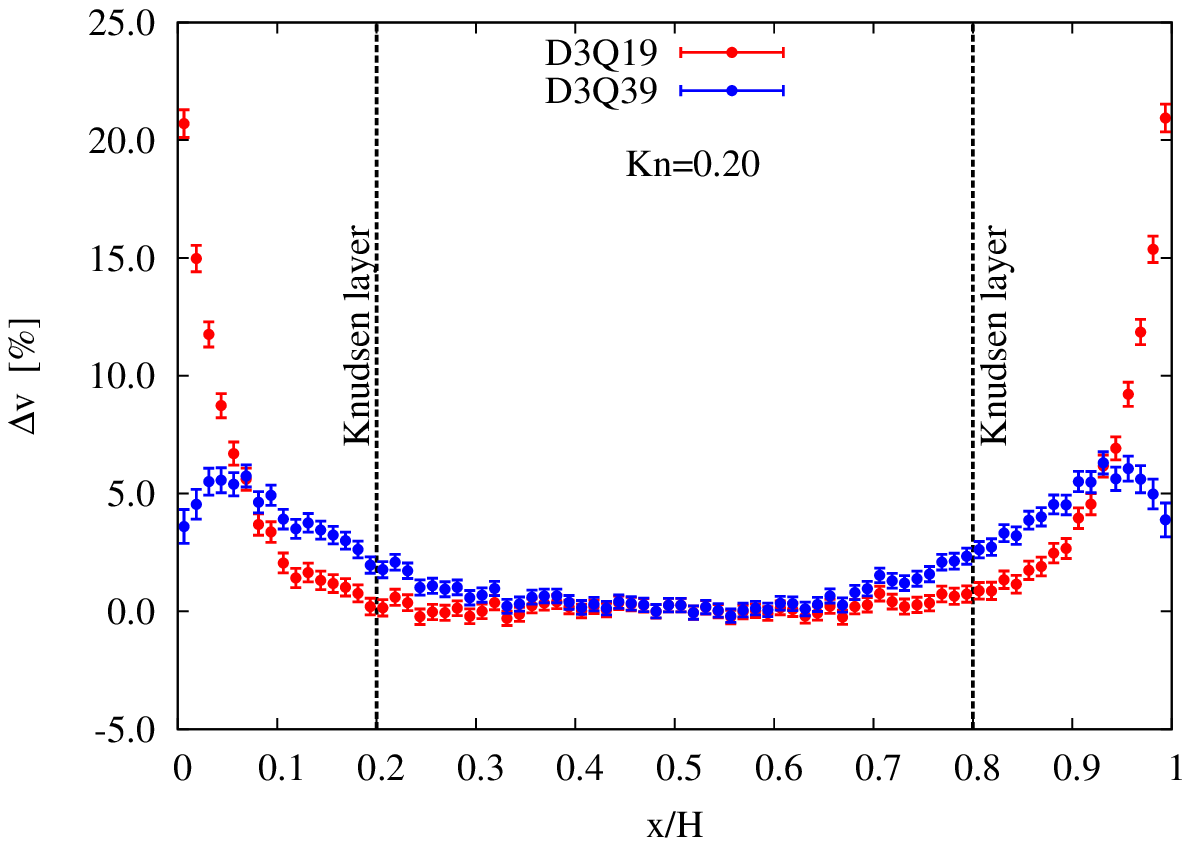}
\includegraphics[width=0.55\textwidth]{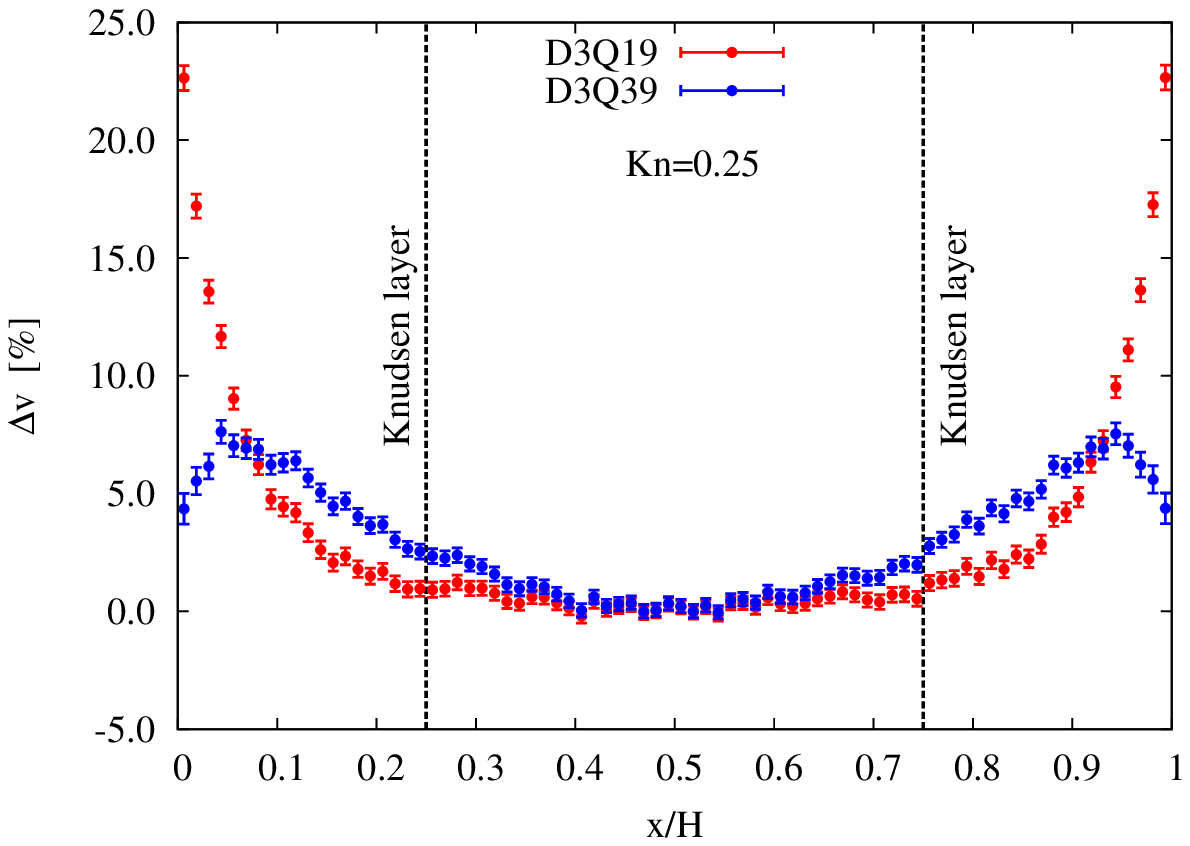}
\caption{Relative error between LBM results and DSMC data, according
to Eq. (\ref{eq:diff}) for Kn=0.10 - 0.25. Dashed vertical lines represent
the boundaries of the Knudsen layer. Error bars from
DSMC simulations on fluid velocity are shown.} \label{fig:profiles_diff}
\end{center}
\end{figure}
In Figure \ref{fig:profiles}, the velocity profiles along the
direction of the forcing, obtained from the
LB models and DSMC, normalized with the centerline velocity, are shown
for Kn=0.15.\\
In the DSMC an \emph{Argon}-like gas has been simulated
and the grid resolution, kept the same for all performed tests, is 
based on the requirements of a DSMC simulation at Kn=0.05.
In all the DSMC simulations, 100 computational molecules are initially
placed in each cell of the domain.\\
LB solution has been considered as converged to the final solution once the following criterion
is fulfilled:
\begin{equation}
\sum_i \frac{\left|\mathbf{u}(\mathbf{x}_i,t)-\mathbf{u}(\mathbf{x}_i,t-1)\right|}{\left|\mathbf{u}(\mathbf{x}_i,t)\right|}< 10^{-6}.\label{eq:convergence}
\end{equation}
In Eq. (\ref{eq:convergence}), $\mathbf{u}(\mathbf{x}_i,t)$ represents
the fluid velocity at the lattice nodes at time $t$. For DSMC, instead,
a 1\% fractional error on fluid velocity components is set as the
requirement to assume the solution as converged; see Section \ref{sec:efficiency}
for the implications in the number of required time steps to achieve
such error.\\
Plots similar to the one of Figure \ref{fig:profiles}, have been drawn also 
for other Kn numbers but they are not reported here. It is more informative, 
in fact, to inspect the relative errors between DSMC and LBM data as done in 
Figure \ref{fig:profiles_diff}. The relative error is defined
as:
\begin{equation}
\Delta v = \frac{v_{LBM} - v_{DSMC}}{v_{LBM}}\label{eq:diff}
\end{equation}
and it is shown for simulations at Kn=0.10-0.25.\\
In the plots of Figure \ref{fig:profiles_diff}, moreover,
the boundaries of the Knudsen layer (black dashed vertical 
lines) are also drawn.
The Knudsen layer is a region in proximity of a solid wall
which extends within the flow domain up to a distance of the 
order of one mean free path. Inside this region 
non-equilibrium effects of the flow are stronger \cite{Gallis2006,Lilley2008}.
It is also within this layer that departures between the numerical
methods become apparent.
In fact, while both LB models provide comparable results in the 
bulk of the flow, within the Knudsen layer they differ
significantly.
The D3Q19 lattice, in fact, recovering up to the Navier-Stokes
level of description only, is rather inaccurate in this part of the
domain.
The D3Q39 lattice, instead, is able to reproduce the DSMC
data to a much better degree of accurary. However, already at Kn=0.25,
it is possible to notice some deviations also within the 
Knudsen layer as the maximum relative error is about equal to 7.5\%.
This behavior can be explained taking into account
that non-equilibrium effects at an order higher than the
third may start to play a role.\\ 
With this statement, we do not imply that LBM is able to reliably simulate rarefied gas flows
only for Kn$\leq$0.25, but that with the current LB model we found reasonable agreement
with DSMC data up to that Kn number. With larger Gauss-Hermite quadratures, in fact,
being able to go beyond the third-order in Hermite polynomials expansion guaranteed
by the D3Q39, further non-equilibrium effects should be correctly captured.
However, we decided not to go further because the next quadrature possessing a high
enough algebraic precision to allow an accurate fourth-order in Hermite polynomials expansion 
involves 91 discrete speeds \cite{Shan2010}.

\subsection{Numerical results for the DSMC2LB mapping scheme}
\begin{figure}
\begin{center}
\includegraphics[width=0.70\textwidth]{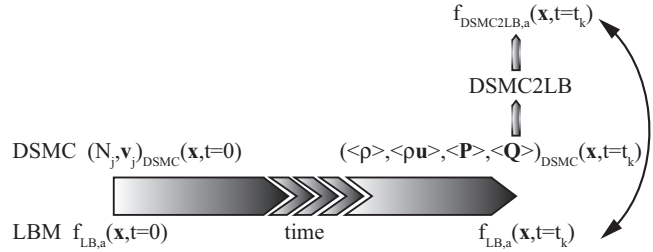}
\caption{Schematic representing the procedure used to compare
the discrete populations built from the DSMC hydrodynamic
moments following the projection DSMC2LB algorithm, $f_{\textrm{DSMC2LB},a}(\mathbf{x},t=t_k)$,
with native discrete populations obtained from an independent
LBM simulation, $f_{\textrm{LB},a}(\mathbf{x},t=t_k)$, under the same flow conditions, 
namely Kn and Ma, at time $t=t_k$, when the steady-state condition
is reached.}\label{fig:timeline_DSMC2LB}
\end{center}
\end{figure}

\begin{figure}
\begin{center}
\includegraphics[width=0.5\textwidth]{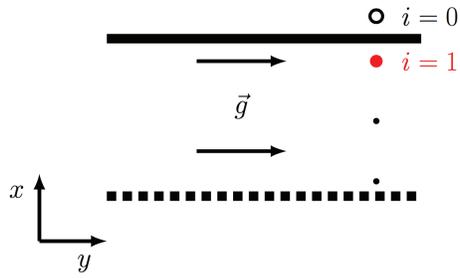}
\caption{Sketch showing the location of the node (red node at 
$i=1$) where data plotted in Figures \ref{fig:dsmc_lbm_015}, 
and \ref{fig:distr_0_15} 
are taken.
$\vec{g}$ represents the body force driving the fluid.}\label{fig:schematic}
\end{center}
\end{figure}

\begin{figure}[ht!]
\begin{center}
\includegraphics[width=0.68\textwidth]{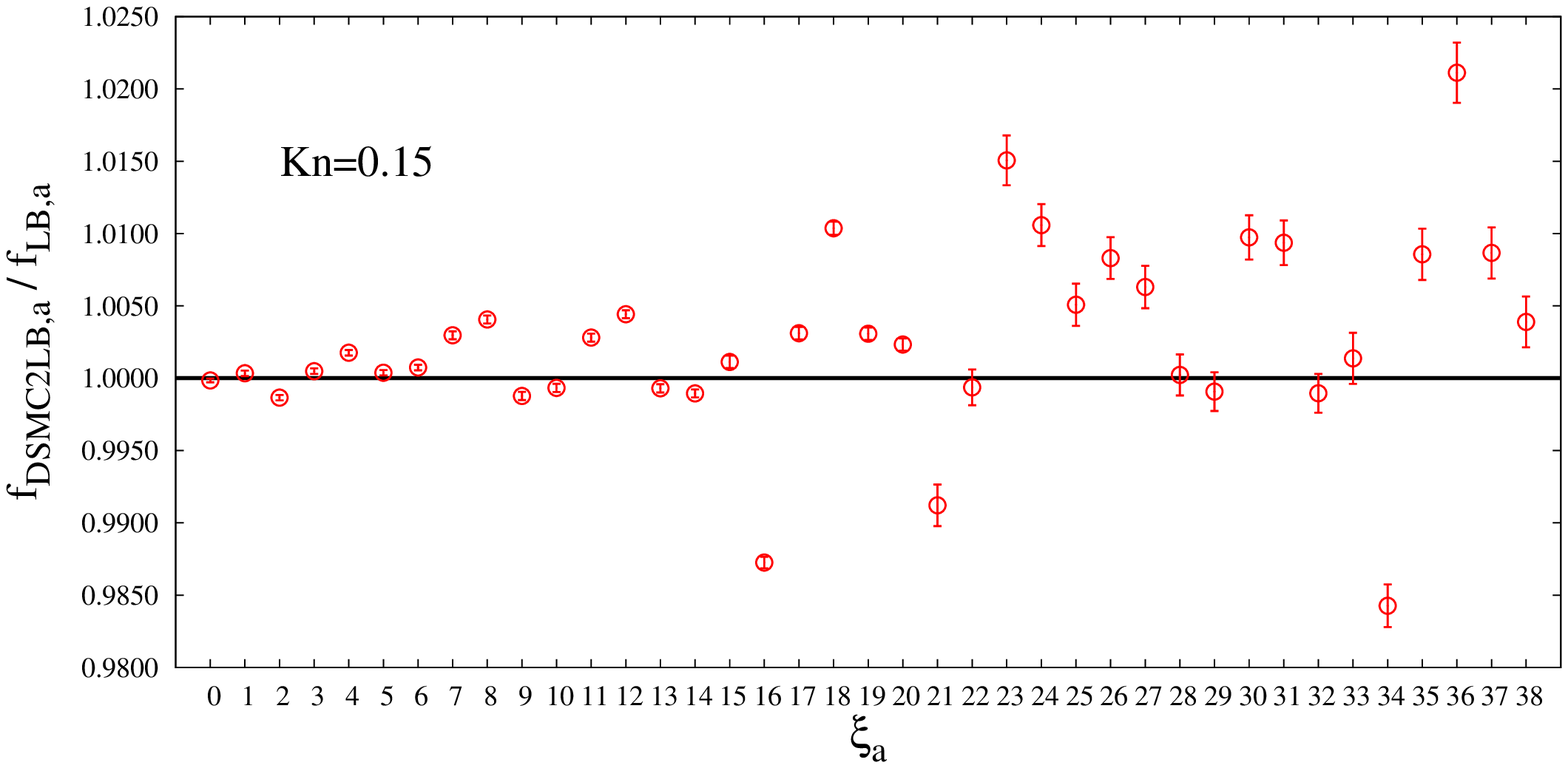}\\
\includegraphics[width=0.68\textwidth]{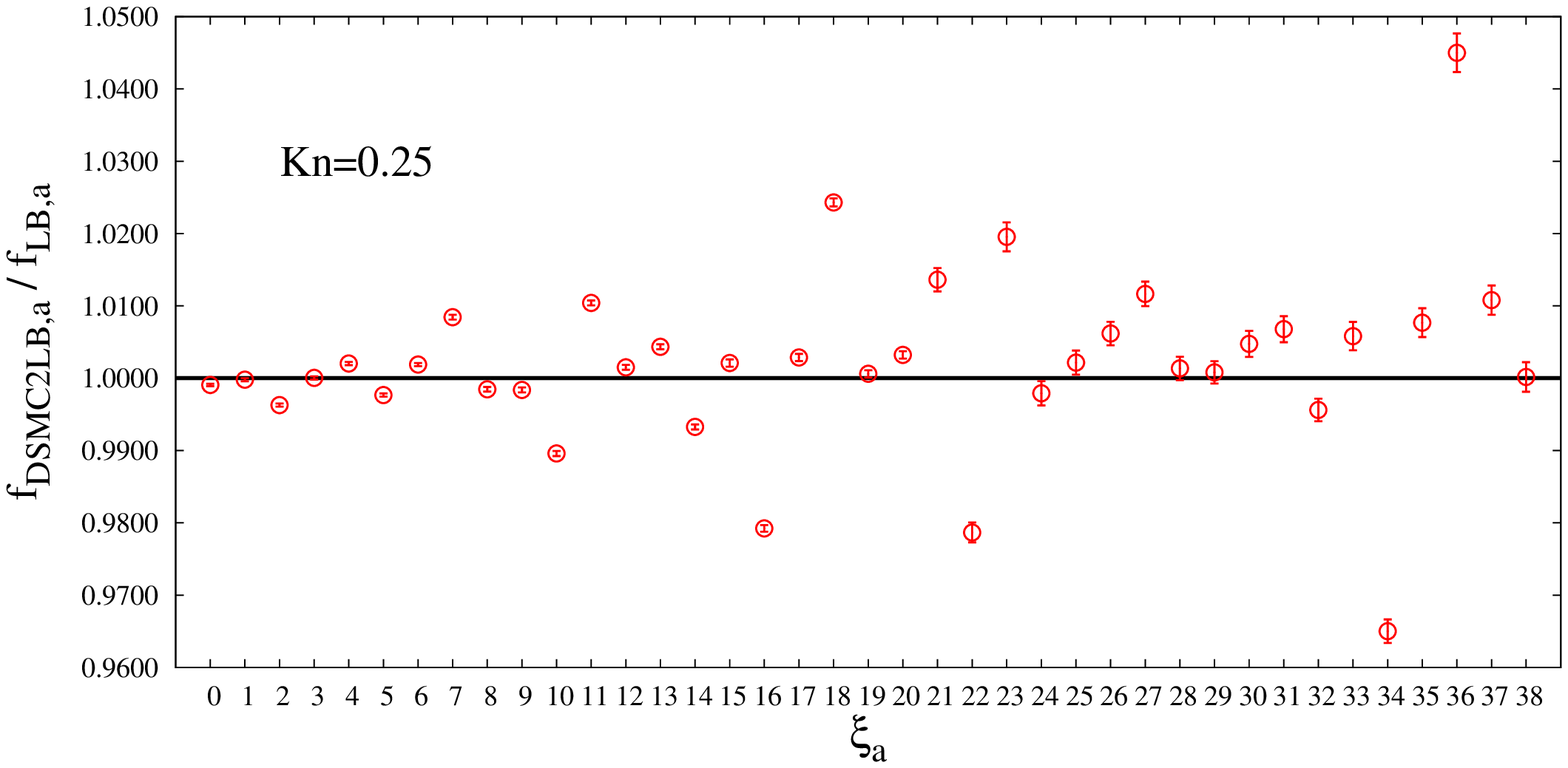}
\caption{Ratio $f_{\textrm{DSMC2LB},a}/f_{\textrm{LB},a}$ where $f_{\textrm{DSMC2LB},a}$ are computed
from Eq. (\ref{eq:discretefunctions}) for plane Poiseuille flow at Kn=0.15 (top) and
at Kn=0.25 (bottom).} \label{fig:dsmc_lbm_015}
\end{center}
\end{figure}

Having concluded that the LBM D3Q39 model provides, for the
problem at hand, a reasonable accurate solution for $Kn \leq 0.25$,
we analyze results related to the mapping scheme step
that allows to project the DSMC hydrodynamic variables
onto the LBM discrete distribution functions for the 
D3Q39 lattice (DSMC2LB projection step).\\
To be noted that the unit conversion as delineated in \ref{sec:appendix}
to pass from SI units, proper of the DSMC method, to the lattice units, 
proper of the LB method, is applied during simulations.
To validate the procedure outlined in Section \ref{sec:dsmc2lb},
we ran two sets of independent DSMC and LBM simulations under
the same force-driven plane Poiseuille flow with Ma based on
the centerline velocity equal to 0.1 and for several Kn numbers.
We verified the accordance between the discrete distributions
functions as computed from the LBM, $f_{\textrm{LB},a}$, and as
obtained from the DSMC2LB projection scheme, $f_{\textrm{DSMC2LB},a}$,
applying Eq. (\ref{eq:discretefunctions}).
In Figure \ref{fig:timeline_DSMC2LB}, a sketch showing the 
procedure to compare the $f_{\textrm{LB},a}$ with the $f_{\textrm{DSMC2LB},a}$
is depicted.
Data refers to the first fluid node/cell in proximity to the wall
located at $x=H$ as shown in the sketch of Figure \ref{fig:schematic}.
In Figure \ref{fig:dsmc_lbm_015} 
%and \ref{fig:dsmc_lbm_025}
the ratio $f_{\textrm{DSMC2LB},a}/f_{\textrm{LB},a}$ is plotted for all discrete speeds 
 $a=0,\dots,d-1$ and for Kn=0.15 and Kn=0.25.
The larger errors that can be detected are about equal to 2\% ($f_{\textrm{DSMC2LB},36}/f_{\textrm{LB},36}\approx1.02$)
and to 5\% ($f_{\textrm{DSMC2LB},36}/f_{\textrm{LB},36}\approx1.05$) for the simulations
at Kn=0.15 and 0.25, respectively. Most of the other ratios are such 
that the error is below 1\%.\\
The error bars present in the plots derive from the fact that we use
the DSMC hydrodynamic moments to build the truncated distributions 
$f^N_{\textrm{DSMC}}(\mathbf{x},\boldsymbol\xi_a,t)$ and those are inherently characterized
by statistical noise.\\
We also note that the larger error bars are present for the discrete
speeds with larger module. This may be attributed to the fact that the magnitude of
the discrete distribution function, $f_a$, is smaller the larger the module
of the corresponding discrete speed, $\boldsymbol\xi_a$, while the statistical noise
does not depend on the particular discrete speed.\\
From the comparison of the discrete distributions, $f_a$, only, however,
it is difficult to understand if the projection mapping scheme is providing accurate
results. So it is more informative to compute the hydrodynamic moments
from $f_{\textrm{LB},a}$ and from $f_{\textrm{DSMC2LB},a}$ at the same node
depicted in Figure \ref{fig:schematic}. The first few moments are reported in
Table \ref{tab:moments}.\\
It can be seen that a good matching is found always within the error bars.\\
Concluding, the projection mapping scheme is able to pass from the DSMC hydrodynamic
quantities to the LBM discrete distributions preserving a reasonable
level of accuracy.
\begin{table*}[ht!]
{\small
\hfill{}
\begin{tabular}{ c c c c }
\hline
\hline
\textbf{Kn=0.15}&$\rho$ [l.u.]& $\rho u_y$ [l.u.] & $P_{xy} + \rho u_xu_y$ [l.u.]\\
\hline
 LBM & 1.0 & 0.0282 & -0.0151 \\
\hline
 DSMC2LB & 1.002 $\pm$ 0.007 & 0.0277 $\pm$ 0.0042 & -0.0146 $\pm$ 0.0027 \\
\thickhline
\textbf{Kn=0.25}&$\rho$ [l.u.]& $\rho u_y$ [l.u.] & $P_{xy} + \rho u_xu_y$ [l.u.]\\
\hline
 LBM & 1.0 & 0.0352 & -0.0180 \\
\hline
 DSMC2LB & 0.999 $\pm$ 0.007 & 0.0340 $\pm$ 0.0042 & -0.0167 $\pm$ 0.0024 \\
\hline
\hline
\end{tabular}}
\hfill{}
\caption{Comparison between the first few moments as computed from $f_{\textrm{DSMC2LB},a}$ 
obtained from the projection mapping scheme and from the native LBM simulations, $f_{\textrm{LB},a}$, at the node 
depicted in Figure \ref{fig:schematic}. Moments are expressed in lattice units.}\label{tab:moments}
\end{table*}
\subsection{Numerical results for the LB2DSMC mapping scheme}
\begin{figure}
\begin{center}
\includegraphics[width=0.65\textwidth]{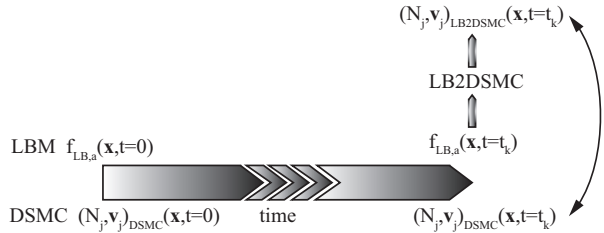}
\caption{Schematic representing the procedure used to compare
the distribution function of the velocity component, $v_{j,\textrm{DSMC}}$, 
obtained from a native DSMC simulation with the distribution function 
of the velocity of the particles, $v_{j,\textrm{LB2DSMC}}$, obtained from the 
reconstruction mapping algorithm LB2DSMC, under the same flow conditions, 
namely Kn and Ma, at time $t=t_k$, when the steady-state condition is reached.}\label{fig:timeline_LB2DSMC}
\end{center}
\end{figure}

\begin{figure}
\begin{center}
\includegraphics[width=0.68\textwidth]{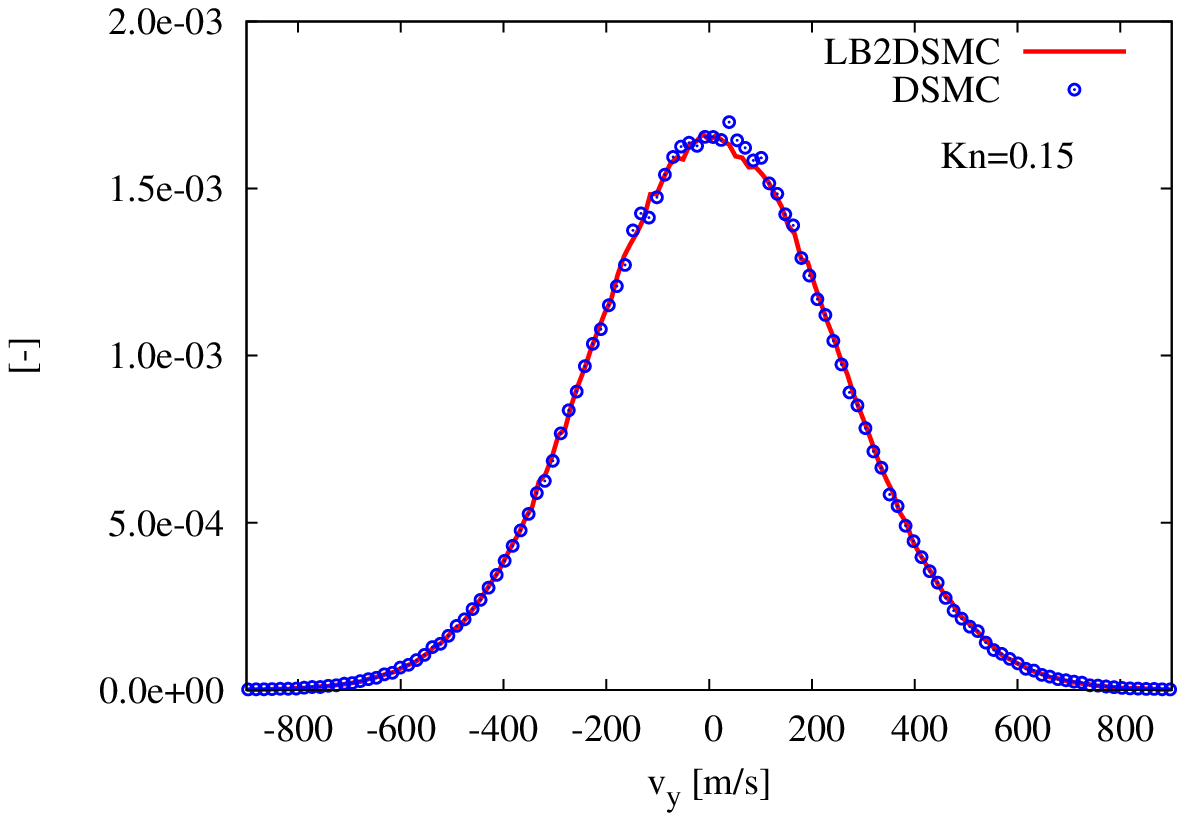}
\includegraphics[width=0.68\textwidth]{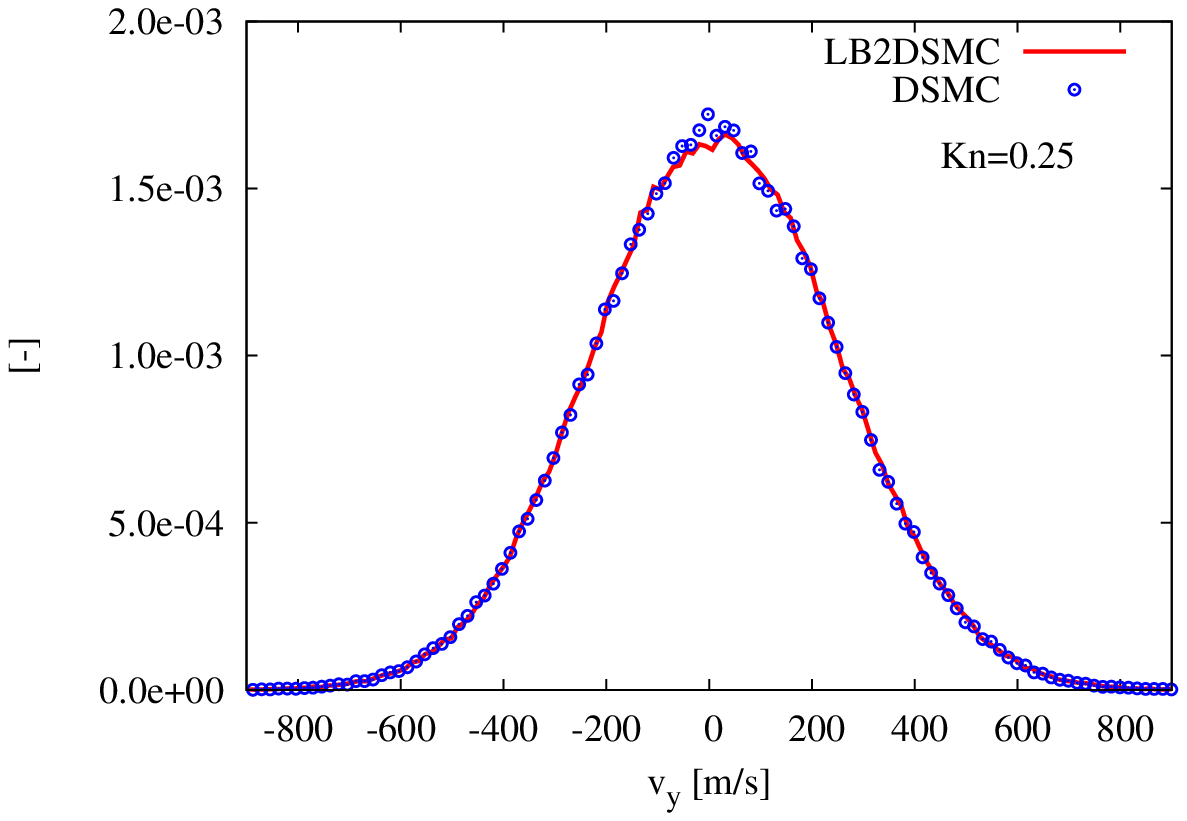}
\caption{Distribution functions for the y-component, $v_y$, of the velocity of the particles,
expressed in DSMC units, as obtained from the native DSMC simulation and from the 
reconstruction mapping scheme using the
algorithm LB2DSMC outlined in Table \ref{tab:sampling} for Kn=0.15 (top)
and Kn=0.25 (bottom), for the cell identified in Figure \ref{fig:schematic}.}\label{fig:distr_0_15}
\end{center}
\end{figure}

\begin{figure}[ht!]
\begin{center}
\includegraphics[width=0.68\textwidth]{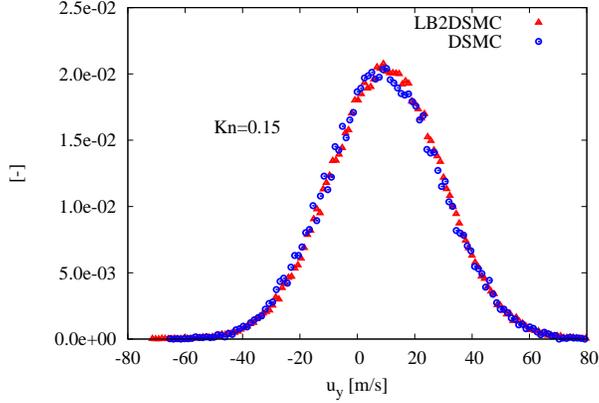}
\caption{Fluid velocity distribution functions for the y-component, $u_y$, expressed in DSMC units,
as obtained from the native DSMC simulation and from the reconstruction mapping scheme using the
algorithm LB2DSMC outlined in Table \ref{tab:sampling} for Kn=0.15, for
the cell identified in Figure \ref{fig:schematic}.}\label{fig:distr_fluid}
\end{center}
\end{figure}

We now move on to analyse the results related to the reconstruction mapping scheme 
step that allows to reconstruct from the LBM discrete distributions, $f_{\textrm{LB},a}$,
the continuous truncated distribution function from which the velocities of 
the DSMC particles can be sampled (LB2DSMC reconstruction step).\\
The unit conversion as delineated in \ref{sec:appendix}
to pass from lattice units, proper of the LB method, to the SI 
units, proper of the DSMC method, is applied during simulations.
As done for the previous step, to validate the procedure outlined in Section
\ref{sec:lb2dsmc}, we ran two independent set of DSMC and LBM 
simulations under the same force-driven plane Poiseuille flow with Ma=0.1 and
for several Kn numbers.\\
As shown in Figure \ref{fig:timeline_LB2DSMC}, we compared the velocity distribution 
functions as obtained from the DSMC simulation collecting the velocities, $\mathbf{v}_{j,\textrm{DSMC}}$, of the 
particles residing at the cell identified in Figure \ref{fig:schematic} and as 
obtained from the velocities of the particles sampled from the velocity distribution function built as in 
Eq. (\ref{eq:sampledistr}) using the algorithm outlined in Table \ref{tab:sampling}, $\mathbf{v}_{j,\textrm{LB2DSMC}}$.\\
In Figure \ref{fig:distr_0_15}, in particular,
the distributions for the velocity component along the direction
of the forcing, $v_y$, are compared for Kn=0.15 and Kn=0.25, respectively.\\
The mean and the standard deviation for the two cases are collected in 
Table \ref{tab:distributions}.
\begin{table}[ht!]
\centering
\begin{tabular}{ c  c  c  }
\hline
\hline
\small \textbf{Kn=0.15} & \small $\langle v_y\rangle$ m/s& \small $\sigma_{v_y}$ m/s\\
\hline
\small DSMC & \small 10.1 & \small 238.6 \\
\hline
\small LB2DSMC & \small 10.5 & \small 239.3 \\
\thickhline
\small \textbf{Kn=0.25} & \small $\langle v_y\rangle$ m/s& \small $\sigma_{v_y}$ m/s\\
\hline
\small DSMC & \small 12.8 & \small 238.8 \\
\hline
\small LB2DSMC & \small 13.4 & \small 239.6 \\
\hline
\hline
\end{tabular}
\caption{Comparison of the means and standard deviations of 
the distributions of Figures \ref{fig:distr_0_15}, expressed
in DSMC units.}\label{tab:distributions}
\end{table}
The velocities of the particles are collected for both cases after a steady-state
condition has been reached.
The deviations between the means, about 4\% for the case at Kn=0.15 and 
about 5\% for the case at Kn=0.25, are in line with the deviations that
are present in Figure \ref{fig:profiles_diff}. The standard deviations
of the two distributions differ for about 0.3\% for both the cases. Related
to this, it has to be recalled that the temperature of Eq. (\ref{eq:step6})
is the reference temperature imposed in the DSMC simulation. The magnitudes
of these standard deviations are compatible with the reference temperature
(T=273 K) and the molecular mass ($m=6.63 \cdot 10^{-26}$ kg) for the gas
used in the DSMC simulation.\\
It is important, however, that also the distributions of the realizations of
the fluid velocity as obtained from DSMC and from the reconstruction mapping scheme match with 
each other.
This is checked computing the fluid velocities as the instantaneous average 
velocity from all the velocities of particles residing in the chosen cell at
regular time intervals (samples are taken once every 50 time steps) so to have uncorrelated samples.
Also in this case, data are collected once the flow has reached a steady-state
condition.\\ 
In Figure \ref{fig:distr_fluid}, the fluid velocity distributions
are plotted for the case Kn=0.15.\\ 
Both the mean and standard deviations of the distributions obtained from the
two methods are in good agreement, demonstrating that the LB2DSMC reconstruction 
step correctly maps the discrete LB distribution functions into the velocities 
of the DSMC particles.
\section{Hybrid model application}\label{sec:hybridapp}
\begin{figure}
\begin{center}
\includegraphics[width=0.6\textwidth]{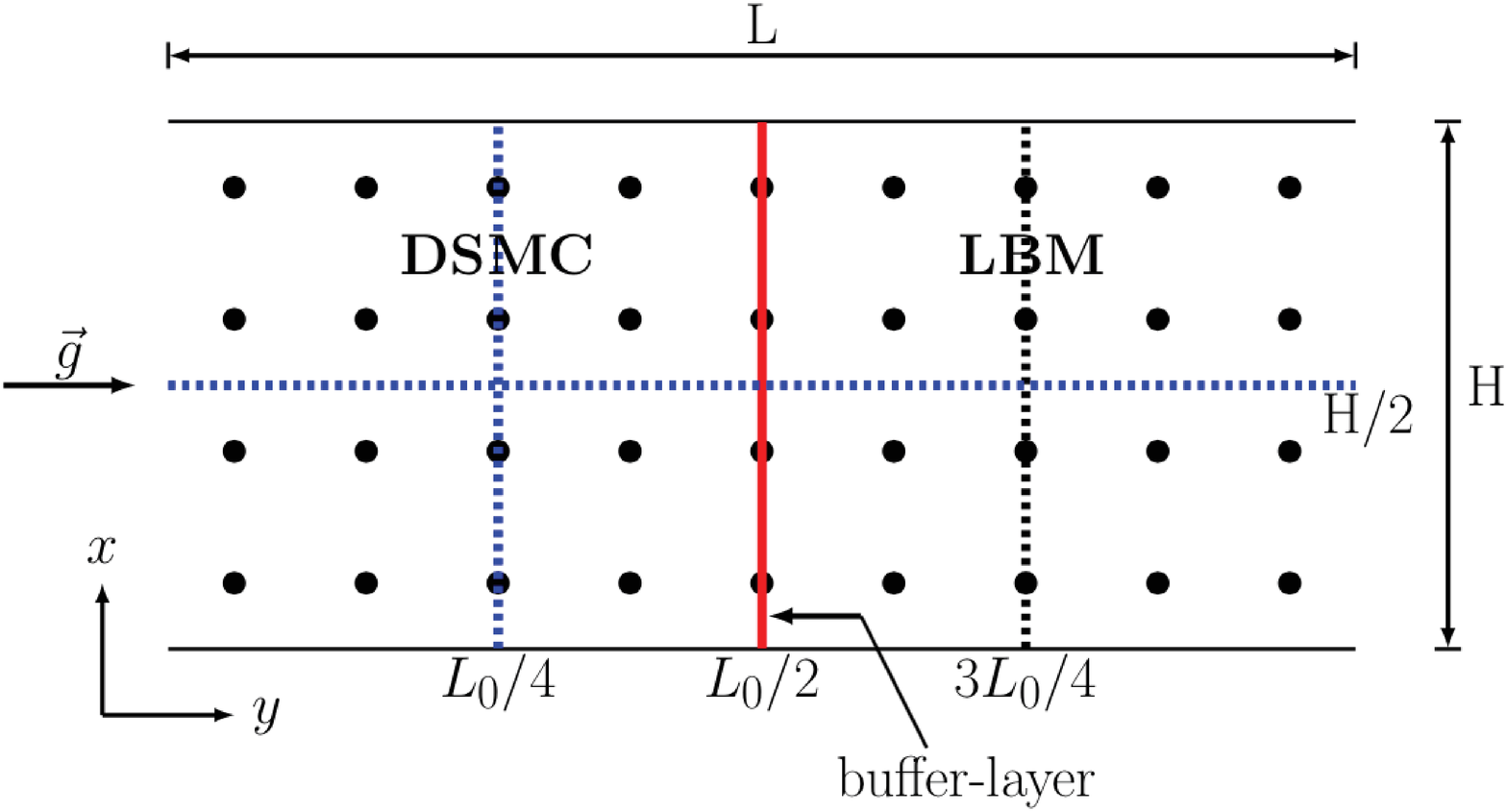}
\caption{Schematic picture of the hybrid model application domain. The
coupling occurs at the LBM lattice sites/DSMC cells placed at $y=L_0/2$. 
The buffer layer comprises only one cell in the streamwise direction 
and all the cells along the transversal direction.
The positions $y=L_0/4$ and $y=3L_0/4$ identify the streamwise
positions where the velocity profiles plotted in Figure \ref{fig:hybrid}
are evaluated.}\label{fig:multiscale}
\end{center}
\end{figure}

\begin{figure}
\begin{center}
\includegraphics[width=0.45\textwidth]{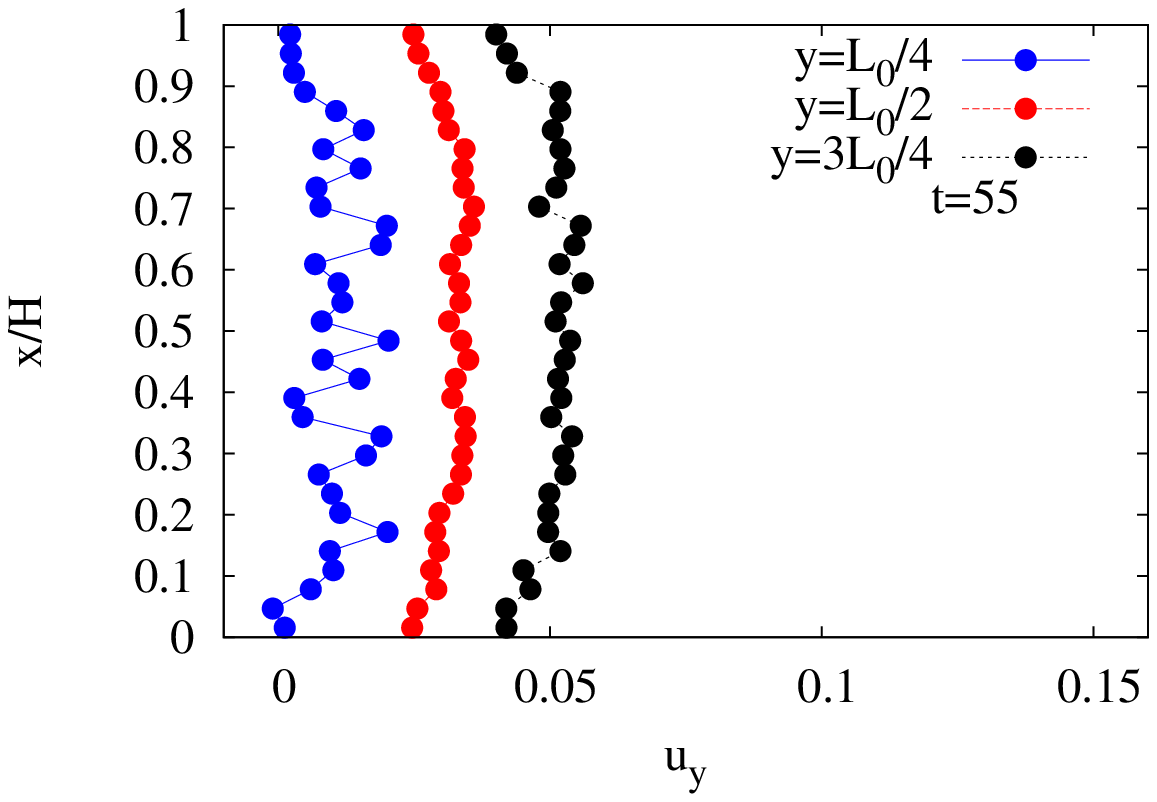}
\includegraphics[width=0.45\textwidth]{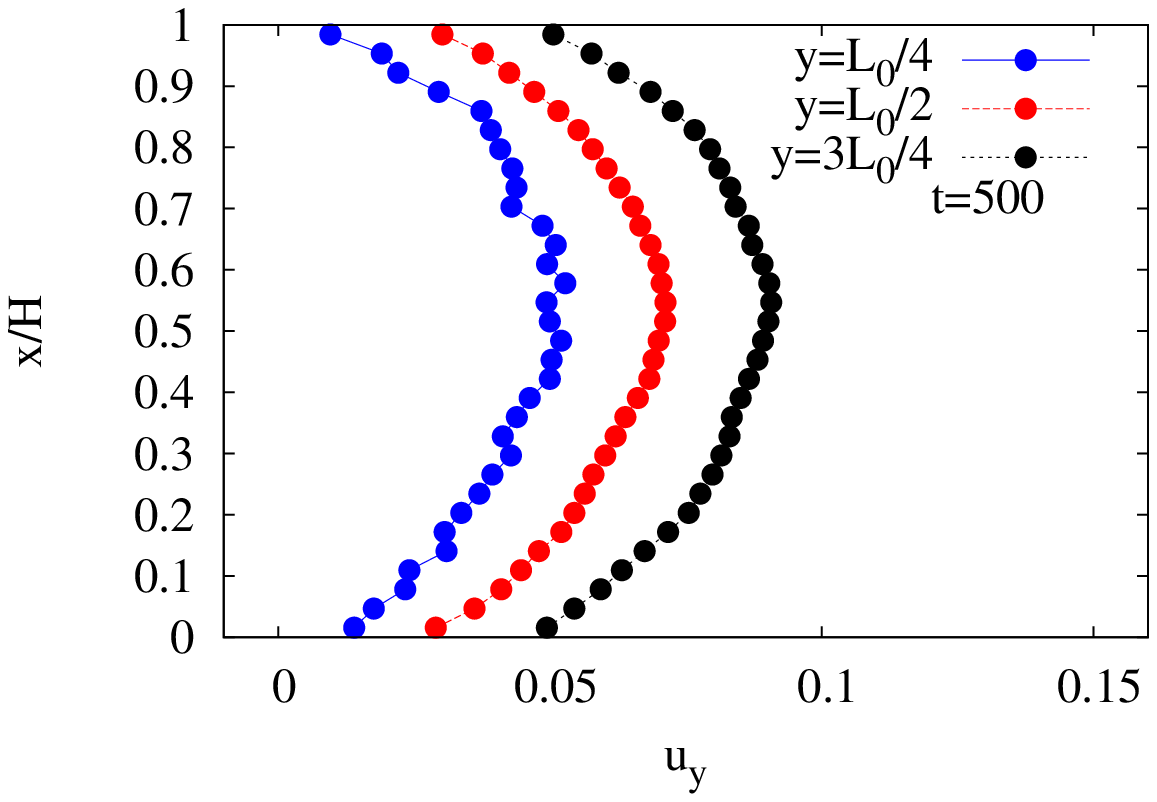}\\
\vspace{-0.3cm}
\includegraphics[width=0.45\textwidth]{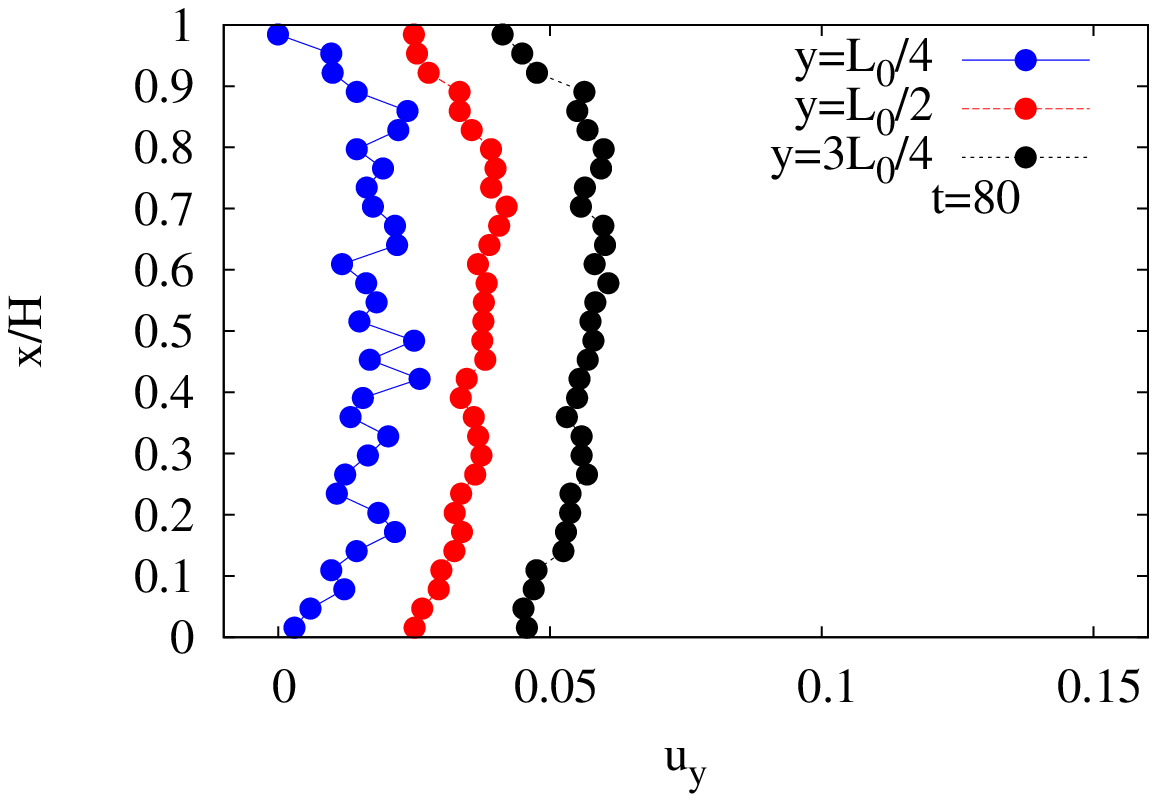}
\includegraphics[width=0.45\textwidth]{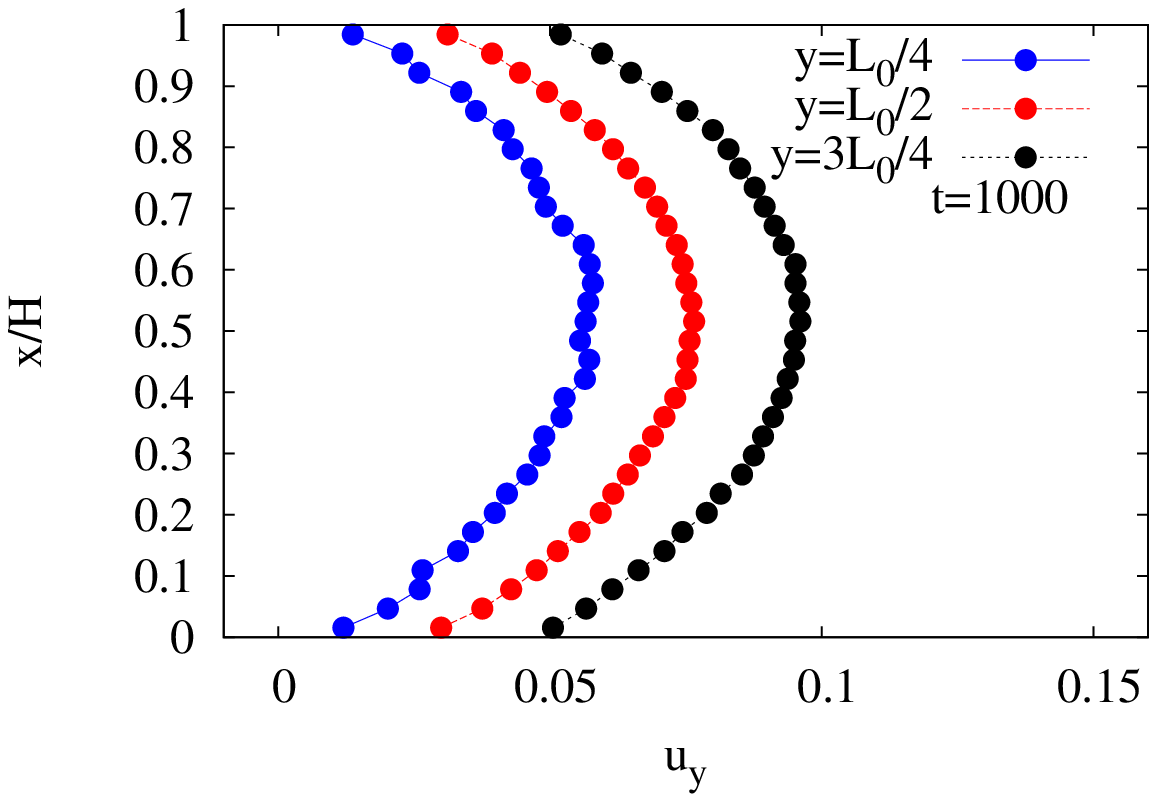}\\
\vspace{-0.3cm}
\includegraphics[width=0.45\textwidth]{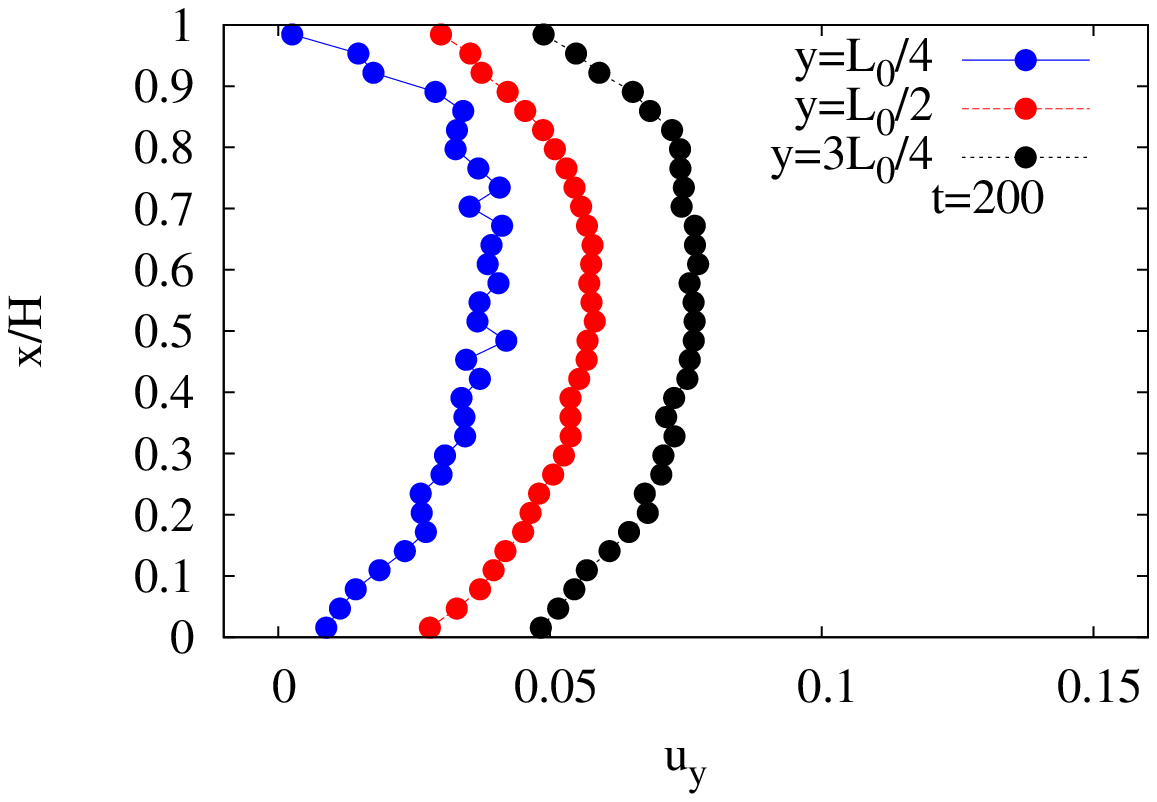}
\includegraphics[width=0.45\textwidth]{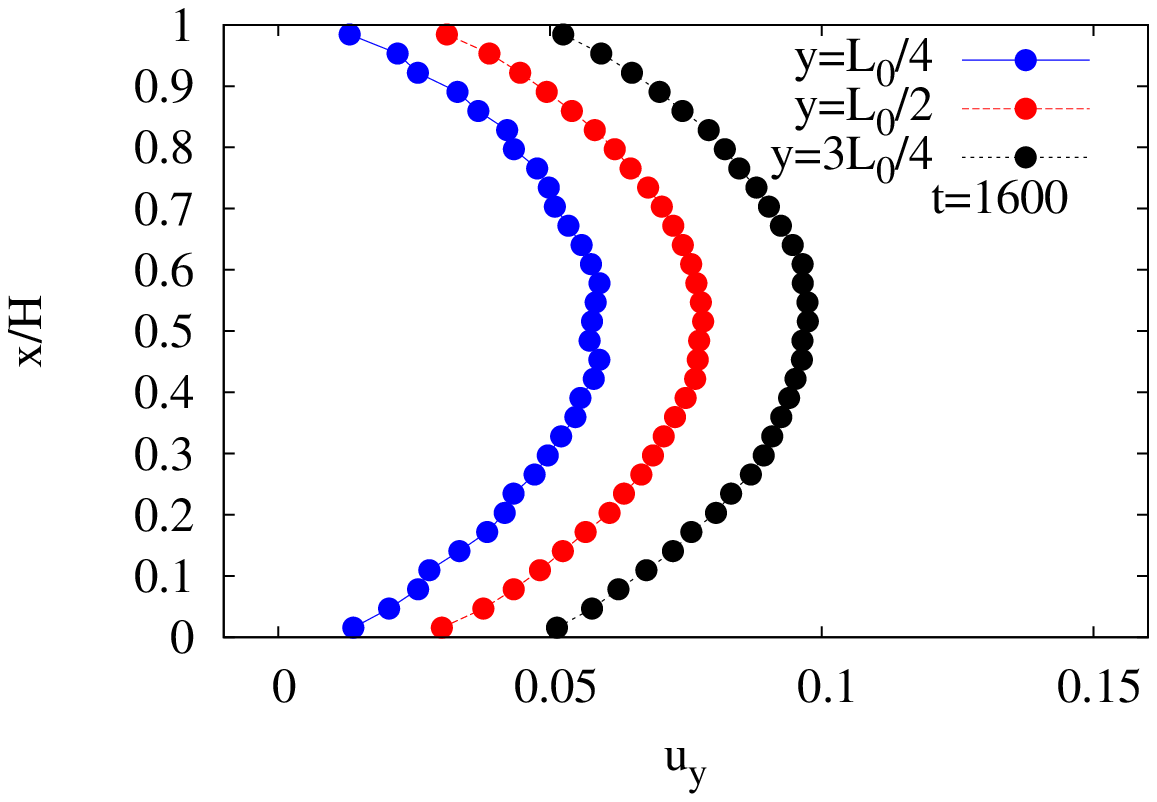}\\
\caption{Evolution in time of the flow when the simplified mapping scheme
DSMC2LB is activated at $y=L_0/2$. Velocity profiles are plotted
in correspondence of the channel sections identified in Figure 
\ref{fig:multiscale}. The velocity profiles, expressed in lattice units, 
at $y=L_0/2$ and $y=3L_0/4$ are shifted +0.02 and +0.04, respectively,
to allow better visualization.} \label{fig:hybrid}
\end{center}
\end{figure}

As a \emph{proof of concept} of a prospective LB-DSMC coupling, 
we applied a hybrid model to a
plane Poiseuille flow with Kn=0.05 and Ma=0.1,
based on centerline velocity.\\
In Figure \ref{fig:multiscale}, the geometry for the application
of the hybrid method is drawn.\\
The domain is divided into two subdomains. In each subdomain,
one solution method is applied. In particular, we assume that, at a section located
at $y=L_0/2$, the two subdomains overlap and this \emph{buffer
layer} is composed by one cell along the flow direction and
extends across the whole height H of the channel.\\
For simplicity, since we wanted to set up the functionality
of the coupling, we use a D3Q19 LB model with kinetic boundary conditions
and no regularization.\\
The mapping scheme, also, is simpler than the one proposed
in Section \ref{sec:remapping}. In particular, we imposed that,
at the centers of DSMC cells/LBM lattice sites within the buffer
layer, the local equilibria are evaluated according to the
hydrodynamic moments computed from the DSMC solution.\\
Operatively, we set the discrete equilibrium distribution
functions, $f^{(0)}_{\textrm{DSMC2LB},a}$, within the buffer layer as:
\begin{equation}
\begin{split}
f^{(0)}_{\textrm{DSMC2LB},a}=w_a \rho_{\textrm{DSMC}} \left[1+\frac{\boldsymbol\xi_a \cdot \mathbf{u_{\textrm{DSMC}}}}{c_s^2} \right.\\
\left. +\frac{\left(\boldsymbol\xi_a \cdot \mathbf{u_{\textrm{DSMC}}} \right)^2}{2c_s^4}-\frac{u_{\textrm{DSMC}}^2}{2c_s^2}\right]\label{eq:feq_dsmc}
\end{split}
\end{equation}
In Figure \ref{fig:hybrid}, we plot the evolution in time of the velocity
profiles obtained from the hybrid method for the test previously introduced.\\
The three plotted profiles represent the data at the three sections along 
the channel located at $y=L_0/4$, $y=L_0/2$ and $y=3L_0/4$. The section
at $y=L_0/4$ is within the DSMC subdomain, while the section at $y=L_0/2$ coincides
with the buffer layer position and the section at $y=3L_0/4$ is within
the LBM subdomain. From the plots of Figure \ref{fig:hybrid}, it is 
possible to see that the inherent statistical noise of the DSMC solution
is \emph{transferred} to the LBM velocity profiles. While averaging
over time, this noise is reduced and also the LBM solution becomes
accordingly, smoothened. Note that, in the DSMC solver, no particular
means to reduce statistical noise, such as variance-reduction methods,
\cite{Kaplan2001,Fan2001,Baker2005}, has been adopted.
Thus, there is certainly room for significant future improvements.\\
From inspection of Figure \ref{fig:hybrid}, it is possible
to detect deviations between the DSMC velocity profile ($y=L_0/4$)
and the LBM velocity profile ($y=3L_0/4$) when the steady state is
reached (see the plot at $t=1600$). These deviations can be attributed to the
limitations of both the LB model and mapping scheme adopted in
this test, as all the non-equilibrium effects have been discarded.\\
The deviations will be removed by adopting the LB model able to extend
the range of applicability of the LBM to rarefied gas flows and by
including non-equilibrium effects in the passage of information 
betweeen the DSMC and the LBM as described in Section \ref{sec:remapping}.
This fully non-equilibrium hybrid model is under development.
\section{Computational efficiency}\label{sec:efficiency}
\begin{figure}
\begin{center}
\includegraphics[width=0.68\textwidth]{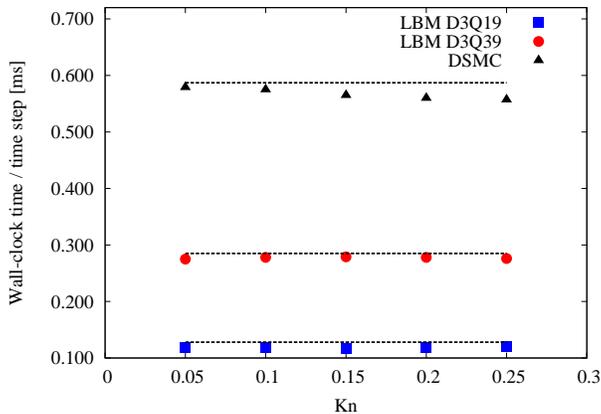}
\caption{Wall-clock time per computational time step in LBM D3Q19, LBM D3Q39
and DSMC for the test of Section \ref{subsec:comparison}.
While LBM data wall clock time does not depend on Kn number, 
DSMC data show a mild dependence on Kn. Note that both LBM and DSMC 
simulations, as stated in Section \ref{subsec:comparison}, are run
on a grid based on the requirements for the DSMC simulation at Kn=0.05
and kept the same for all the simulations at different Kn number.
32000 particles are employed for the DSMC simulations.}\label{fig:wall_clock_time}
\end{center}
\end{figure}

\begin{figure}
\begin{center}
\includegraphics[width=0.68\textwidth]{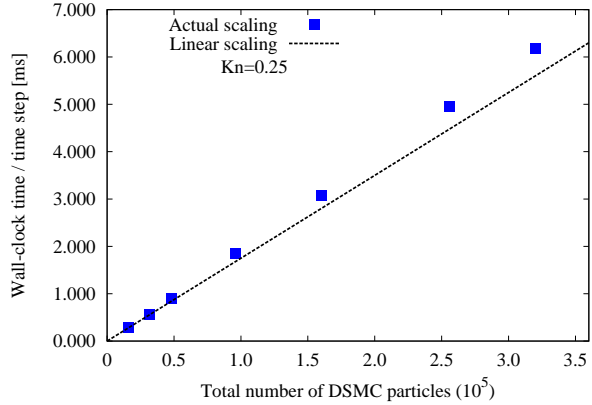}
\caption{Wall-clock time per time step for DSMC as a function of the total number of
particles for a simulation at Kn=0.25. Linear scaling is shown for the range of
particles per cell typically employed in DSMC simulations. Savings in
the total number of particles reflect in a linear reduction of the total 
wall-clock time of the simulation. The simulated flow is the same presented
in Section \ref{subsec:comparison}.}\label{fig:wall_clock_time_dsmc}
\end{center}
\end{figure}

We conclude by comparing the computational efficiency of the two methods
and by estimating the computation times of the hybrid method with respect
to a full DSMC simulation.\\
Both the DSMC and LBM codes are parallelized. All data presented in this 
section are from simulations run on a dual-core PC (Intel Core i5-6300U 2.4 GHz) 
with hyperthreading enabled and refer to the test cases of Sections \ref{subsec:comparison} and 
\ref{sec:hybridapp}.\\
In Figure \ref{fig:wall_clock_time}, the wall-clock time per computational
time step is plotted as a function of the flow Kn number for both LBM models
(D3Q19 and D3Q39 with the regularization procedure and kinetic boundary 
conditions) and DSMC. The wall-clock time per time step is constant for the
LBM simulations while it shows a dependence on Kn for the DSMC method.\\
This might be explained considering that in DSMC, as kinetic theory prescribes,
the total number of interparticle collisions scales with the 
number density. For the NoTimeCounter (NTC) algorithm, \cite{Bird1994}, as the 
one employed here, one has:
\begin{equation}
N_{\textrm{collis}}(t_k,x_j)=\frac{1}{2}N\bar{N}F_N (\sigma_T c_r)_{\max} \Delta t/V_c\label{eq:collisionnumber}
\end{equation}
where $N=n(t_k,x_j)V_c/F_N$, with $n$ number density at time $t_k$ and cell $j$,
$V_c$ the cell volume and $F_N$ the number of real molecules represented
by a simulated computational particle, $\bar{N}=\langle N \rangle$, $\sigma_T$
the molecular cross section, $c_r$ the relative velocity between the selected particles
to undergo collision, $\Delta t$ is the time step duration. 
In Eq. (\ref{eq:collisionnumber}), the term $(\sigma_T c_r)_{\max}$
is the maximum value of the product between the collision cross section and the
relative velocity between the selected particles in each grid cell.\\
From Eq. (\ref{eq:collisionnumber}) the larger the Kn number, the smaller
the number density $n$ and the smaller the total number of collisions.
However, since the same number of cells and particles are used for all
the simulations and since the collision step in the DSMC method is
just one part of the algorithm, only a small decrease in the wall-clock
time of the single computational time step is achieved while increasing 
Kn number.
In the LBM, instead, Kn number determines the relaxation time $\tau$ 
but different values of $\tau$ do not affect
the computational efficiency of the single computational step.\\
From Figure \ref{fig:wall_clock_time}, it is also evident the fact that LBM
wall-clock times are smaller than the ones for DSMC. In particular, a single
computational time step for the D3Q19 model is 5 times faster and for the
D3Q39 model is 2 times faster than for the DSMC.\\
These numbers, however, do not tell the full story because DSMC is intrinsically
characterized by statistical noise due to thermal fluctuations. This greatly 
affects the computational efficiency of the DSMC in comparison with LBM.\\
In fact, to reduce the statistical noise on DSMC hydrodynamic moments, time (or
ensemble) averaging is needed.
For example, one standard deviation on the fluid velocity components measurement,
$\sigma_{u_i}$, is given (at equilibrium) by \cite{Hadji2003}:
\begin{equation}
\sigma_{u_i}=\sqrt{\frac{k_B \langle T \rangle}{m \langle N \rangle}} \frac{1}{\sqrt{S}}
\end{equation}
where $\langle T \rangle$ and $\langle N \rangle$ are the averages of temperature
and number of computational particles in a cell and S is the number of independent
statistical samples.\\
An estimate on the statistical error on the evaluation of the fluid velocity 
is given as:
\begin{equation}
\begin{split}
E_{u_i}=\frac{\sigma_{u_i}}{\left| \langle u_i \rangle \right|}=\sqrt{\frac{k_B \langle T \rangle}{m \langle N \rangle}} \frac{1}{\sqrt{S}} \frac{1}{\left| \langle u_i \rangle \right|}\\
=\frac{1}{\sqrt {\gamma \langle N \rangle S} \:\textrm{Ma}}
\end{split}
\end{equation}
where $\gamma$ is the gas specific heat ratio (1.67 for \emph{Argon}) and Ma
is the Mach number.\\
If a 1\% fractional error is desired, for a Ma=0.1 flow and $\langle N \rangle=100$,
$S \approx 3600$ independent samples are needed. Generally, to obtain independent
samples 10-100 time steps between the samples are required. In all the simulations
in this work, we decided to perform the sampling every 50 time steps. Calculation of
the correlation coefficients between sampled quantities showed that, for the flow
of these tests, a 50 time steps interval is sufficient, e.g.:
\begin{equation}
\textrm{corr}(u_x,u_y)=\frac{\langle \delta u_x \delta u_y \rangle}{\sqrt{\langle \delta u_x^2 \rangle \langle \delta u_y^2 \rangle}}=-0.008.
\end{equation}
Estimates on the number of needed independent samples to reach a given fractional
error and on the size of the time steps interval so to obtain independent samples
allow to determine the number of the required total computational time steps. So,
for the tests we performed, at least 180000 time steps are needed.\\
For the LBM, instead, a steady-state solution is reached in few thousands time steps.
About 5000 time steps are sufficient to reach the final solution.\\
These numbers directly reflect in the comparison between the total wall-clock times
needed for the DSMC and LBM simulations. Using data collected in Figure \ref{fig:wall_clock_time},
if the DSMC is compared with the D3Q19 model, then the latter is about 180 times
faster, while if the comparison is made with the D3Q39 model, then the latter
is about 70 times faster. Moreover, a reduction in the total number of DSMC
particles guaranteed by reducing the domain assigned to the DSMC reflects in a reduction 
in the wall-clock time per time step as shown in Figure \ref{fig:wall_clock_time_dsmc}
where a linear scaling is found for the range of particles typically employed
for the flow under consideration.\\ 
Finally, these numbers allow to estimate the potential gain in efficiency that
can be obtained by the application of the hybrid model.\\
Using the simplified mapping scheme to pass from DSMC to LBM as described in
Section \ref{sec:hybridapp} and assuming that the domain is divided into two 
subdomains of equal size, then a speed-up of about 1.7 with respect 
to a full DSMC simulation over the whole domain is reachable for the tested
Poiseuille flow. To be noted that the over-head due to the application of the 
simplified mapping scheme is very limited since the buffer layer is composed
of just one layer of cells/lattice nodes. For more complicated flows, however, coupling
may be required to be applied over larger overlapping zones.

\section{Conclusions}
We developed a kinetic mapping scheme based on Grad's moments method and
Gauss-Hermite quadrature in view of coupling DSMC and LBM models to simulate
isothermal flows with non-uniform rarefaction effects.
The main steps of the mapping algorithm between DSMC and LBM in order 
to allow an accurate passage between the two methods
domains were discussed. To extend the range of applicability of LBM beyond
the Navier-Stokes equation level, and thus postponing the passage to the DSMC
solver, the need for adopting a high-order lattice (D3Q39) and a regularization
procedure for the LBM is demonstrated by finding a good agreement between
the DSMC and LBM velocity profile for plane Poiseuille flow up to Kn=0.25.
As a proof of concept of the hybrid method, a simpler version of the mapping
scheme which enforces the passage through local equilibrium states has been performed
for the simulation of a plane Poiseuille flow at Kn=0.05.
We have also estimated that the adoption of the hybrid scheme significantly
increases computational efficiency with respect to a DSMC simulation performed
over the whole domain by a factor equal to 1.7 for
the flow conditions shown in the test case. 
The adoption within the hybrid model of the complete mapping scheme 
including non-equilibrium effects is currently under development.

\section*{Acknowledgements}
This research is supported by the Dutch Technology Foundation STW, which is part of the 
Netherlands Organisation for Scientific Research (NWO), and which is partly funded 
by the Ministry of Economic Affairs.

\appendix
\section{Scaling factors}\label{sec:appendix}
To be able to apply the proposed methods also in engineering contexts and in
parallel with experiments, we decided to employ in the DSMC simulations dimensional
variables with SI units. This implies that, prior to any transfer of information
between LB and DSMC, a proper conversion from lattice units to SI units,
or vice versa according to the fact that the DSMC2LB or LB2DSMC mapping scheme is involved, 
has to be performed.\\
The basic elementary conversion scales are here introduced:
\begin{itemize}
\item[-] \emph{Length scale}\\ Since in LB we assume the lattice spacing $\Delta x$
as the space unit and since we impose that the centers of the DSMC cells overlap
with the LB sites, then the length scale is set as:
\begin{equation}
L_0 = \Delta x_{\textrm{DSMC}}\:\:\:\:\textrm{[m]},
\end{equation}
where $\Delta x_{\textrm{DSMC}}$ is the linear distance between the centers of two
adjacent DSMC cells. Note that this implies that, at least in the buffer layer, the
DSMC cells are cubic;
\item[-] \emph{Time scale}\\ Similarly, the time unit within the LB simulation is 
the elementary lattice time-step. The physical value can be defined through the 
speed of sound within the lattice, $c_s$, and of the gas in the DSMC simulation, $a$,
as
\begin{equation}
T_0 = \frac{c_s}{a}\Delta x_{\textrm{DSMC}} \:\:\:\:\textrm{[s]}.
\end{equation}
\item[-] \emph{Mass scale}\\ As the mass within the DSMC cells/LB nodes where 
coupling occurs must be conserved, and assuming the lattice particles are given
a unit mass, then the mass scale can be defined as follows:
\begin{equation}
M_0 = \frac{F_{N,\textrm{DSMC}}\:N_{\textrm{DSMC}}\:m}{\sum_a f_{\textrm{LB},a}} \:\:\:\:\textrm{[kg]},
\end{equation}
where $F_{N,\textrm{DSMC}}$ is the number of real molecules represented by one DSMC
particle, $N_{\textrm{DSMC}}$ is the number of DSMC particles in a cell, and $m$
is the gas molecular mass.
\end{itemize}
From these three scaling factors, it is possible to derive all the other physical conversion scales.

\section*{References}

\bibliography{bibliografia}

\end{document}